\documentclass[aos,MSNbibl,dvips]{arximspdf}
\usepackage{graphicx}
\doi{10.1214/13-AOS1144} %kopijuoti is PTS
\volume{41}
\issue{6}
\pubyear{2013}
\firstpage{2703}
\lastpage{2738}

\makeatletter
\newcommand{\widebar}{\overline}
\newcommand{\rrvert}{\vert}
\newcommand{\llvert}{\vert}
\newproclaim{prob}{Problem}
\newtheorem{theorem}{Theorem}
\newtheorem{lemma}{Lemma}
\newtheorem{proposition}{Proposition}
\newtheorem{corollary}{Corollary}
\newproclaim{assumption}{Assumption}
\newproclaim{ass}{Assumption}
\newproclaim{fact}{Fact}
\newproclaim{definition} {Definition}
\newproclaim{remarks}{Remark}
\newtheorem{claim}{Claim}
\newcommand{\argmax}{\arg\max}
\newcommand{\argmin}{\arg\min}
\makeatother

\begin{document}
\begin{frontmatter}

\title{Active sequential hypothesis testing\thanksref{T1}}
\runtitle{Active sequential hypothesis testing}

\begin{aug}
\author[A]{\fnms{Mohammad} \snm{Naghshvar}\corref{}\ead[label=e1]{mnaghshvar@qti.qualcomm.com}\ead[label=e11]{m.naghshvar@gmail.com}}
\and
\author[B]{\fnms{Tara} \snm{Javidi}\ead[label=e2]{tjavidi@ucsd.edu}}
\runauthor{M. Naghshvar and T. Javidi}
\affiliation{Qualcomm Inc. and University of California, San Diego}
\address[A]{Qualcomm Technologies, Inc.\\
Corporate R\&D\\
5775 Morehouse Drive\\
San Diego, California 92121\\
USA\\
\printead{e1}} %adresu isvedimo komanda gale!
\address[B]{Department of Electrical\\
\quad and Computer Engineering\\
University of California, San Diego\\
La Jolla, California 92093\\
USA\\
\printead{e2}}
\end{aug}
\thankstext{T1}{Supported in part by the industrial sponsors of UCSD
Center for Wireless Communication (CWC)
and Center for Networked Systems (CNS), and NSF Grants CCF-1018722 and
AST-1247995.}

% HISTORY:
\received{\smonth{4} \syear{2013}}

% ABSTRACT
%
\begin{abstract}
Consider a decision maker who is responsible to dynamically collect
observations so as to enhance his information about an underlying
phenomena of interest in a speedy manner while accounting for the
penalty of wrong declaration.
%
%The special cases of the problem are shown to be
%that of noisy dynamic search and variable-length coding with feedback.
Due to the sequential nature of the problem, the decision maker relies
on his current information state to adaptively select the most
``informative'' sensing action among the available ones.

In this paper, using results in dynamic programming, lower bounds for
the optimal total cost are established. The lower bounds characterize
the fundamental limits on the maximum achievable information
acquisition rate and the optimal reliability.
%(the expected total number of samples plus the penalty of wrong
%declaration)
Moreover, upper bounds are obtained via an analysis of two heuristic
policies for dynamic selection of actions. It is shown that the first
proposed heuristic achieves asymptotic optimality, where the notion of
asymptotic optimality, due to Chernoff, implies that the relative
difference between the total cost achieved by the proposed policy and
the optimal total cost approaches zero as the penalty of wrong
declaration (hence the number of collected samples) increases.
%Furthermore, the gain of adaptive selection of sensing actions is
%shown to be
%at least logarithmic in the penalty associated with wrong declarations.
%
The second heuristic is shown to achieve asymptotic optimality only in
a limited setting such as the problem of a noisy dynamic search. % and
%variable-length coding with feedback.
However, by considering the dependency on the number of hypotheses,
under a technical condition, this second heuristic is shown to achieve
a nonzero information acquisition rate, establishing a lower bound for
the maximum achievable rate and error exponent.
In the case of a noisy dynamic search with size-independent noise, %and
%variable-length coding with feedback,
the obtained nonzero rate and error exponent are shown to be maximum.
\end{abstract}

% KEYWORDS

\end{frontmatter}

\newpage
\setcounter{footnote}{1}
%s1 #&#
\section{Introduction}\label{sec1}

This paper considers a generalization of the classical sequential
hypothesis testing problem due to Wald~\cite{Wald48}. Suppose there are
$M$ hypotheses among which only one is true. A Bayesian decision maker
is responsible to enhance his information about the correct hypothesis
in a speedy and sequential manner while accounting for the penalty of
wrong declaration. In contrast to the classical sequential $M$-ary
hypothesis testing problem~\cite{Armitage50,Lorden77,Dragalin99}, our
decision maker can choose one of $K$ available actions and, hence,
exert some control over the collected samples' ``information content.''
We refer to this generalization, originally tackled by
Chernoff~\cite{Chernoff59}, as the \textit{active} sequential
hypothesis testing problem.

%%in cognition, communications, design of experiments, and sensor
%management.
%For instance, consider the classic ``where is Waldo?'' game
%in which a child is given an image and is asked to identify the
%location of Waldo against
%a crowded background. The child's gaze not only provides samples which
%successively enhance the child's belief about Waldo's location,
%but its focus on a particular segment also controls the ``information
%content'' of the collected samples.
%Similar conceptual problem arises in}
The active sequential hypothesis testing problem naturally arises in a
broad spectrum of applications such as medical
diagnosis~\cite{Berry11}, cognition~\cite{Shenoy11}, sensor
management~\cite{Hero11}, underwater inspection~\cite{Hollinger11},
generalized search~\cite{Nowak11IT}, group testing~\cite{Chan11} and
channel coding with perfect feedback~\cite{Burnashev76}. It is
intuitive that at any time instant, an optimized Bayesian decision
maker relies on his current belief to adaptively select the most
``informative'' sensing action, that is, an action that provides the
highest amount of ``information.''
%and \mbox{(re-)evaluate} the trade-off between the observation
%precision and the cost of various actions.
Making this intuition precise is the topic of our study.

The most well-known instance of our problem is the case of binary
hypothesis testing with passive sensing ($M=2$, $K=1$), first studied
by Wald~\cite{Wald48}. In this instance of the problem, the optimal
action at any given time is provided by a sequential probability ratio
test (SPRT).
%This simple ratio test fails to be optimal when $M>2$ and $K=1$. In
%fact, the analytical and precise characterization of the optimal
%policy in this case remains an open problem, despite
There are numerous studies on the generalizations to $M>2$ ($K=1$) and
the performance of known simple and practical heuristic tests such as
MSPRT~\cite{Armitage50,Lorden77,Dragalin99}. The generalization to the
active testing case was considered by Chernoff in~\cite{Chernoff59}
where a heuristic randomized policy was proposed and whose asymptotic
performance was analyzed. More specifically, under a certain technical
assumption on uniformly distinguishable hypotheses, the proposed
heuristic policy is shown to achieve asymptotic optimality where the
notion of asymptotic optimality~\cite{Chernoff59} denotes the relative
tightness of the performance upper bound associated with the proposed
policy and the lower bound associated with the optimal policy.

The problem of active hypothesis testing also generalizes another
classic problem in the literature: the comparison of experiments first
introduced by \mbox{Blackwell}~\cite{Blackwell53}. This is a single-shot
version of the active hypothesis testing problem in which the decision
maker can choose one of several (usually two) actions/experiments to
collect a single observation sample before making the final decision.
%However, the decision maker has access to several (usually two)
%actions/experiments.
There have been extensive
studies~\cite{Blackwell53,Lindley56,Lecam64,DeGroot70,Goel79,Lehmann88,Torgersen91}
on comparing the actions.
% and finding conditions under which one action is ``better'' than
%others, i.e., by using the observations from a particular sensing
%action, the decision maker can do at least as well as by using the
%observations from other actions.
%Despite its long history in the stochastic control literature,
%however, the analytic characterization of the solution
%is still not well understood.
%
Applying various results from~\cite{Blackwell53,DeGroot70} in our
context of active hypothesis testing and utilizing a dynamic
programming interpretation, a notion of optimal information utility,
that is, an optimal measure to quantify the information gained by
different sensing actions, can be derived~\cite{ISIT10}.
%Inspired by this view of the problem, which coincides with that
%promoted by DeGroot \cite{DeGroot62},
%we provide a set of (uniform) lower bounds for optimal information
%utility and a set of corresponding
%active sensing heuristics. The construction of these heuristics relate
%various notions of information utility with the statistical properties
%of the outcome such as Kullback--Leibler divergence and mutual
%information.
%Via an asymptotic analysis, the performance of these policies are
%investigated. More specifically, we provide two
%heuristic policies.
Inspired by this view of the problem, which coincides with that
promoted by DeGroot~\cite{DeGroot62}, we provide a set of (uniform)
lower bounds for the optimal information utility. Furthermore, we
provide two heuristic policies whose performance is investigated via
nonasymptotic and asymptotic analysis. %%
The first policy is shown to be asymptotically optimal, matching the
performance of the scheme proposed in~\cite{Chernoff59}
(and follow-up works~\cite{Bessler60,Blot73}), %~\cite{Blot73,%NitinawaratArxiv}),
and provides a benchmark for comparison when considering Chernoff's
asymptotic regime.
%while relaxing the technical assumption on uniform distinguishibiliy
%of the hypotheses.
%In contrast to policies proposed in
%is
In contrast, our second proposed policy is only shown to be
asymptotically optimal in a limited setup, including that of noisy
dynamic search.
However, this policy has a provable advantage for large $M$ over
%our first policy, as well as other solutions in the literature.
those proposed in the literature. More specifically, this policy can
provide, under a technical condition, reliability and speedy
declaration simultaneously. In information theoretic terms, this policy
can be shown to achieve nonzero information acquisition rate and,
hence, to generalize Burnashev's~\cite{Burnashev76} variable-length
channel coding scheme. We elaborate on a complete literature survey in
Section~\ref{sum}.

The remainder of this paper is organized as follows. In
Section~\ref{PFDP} we formulate the active sequential hypothesis
testing problem and discuss the related works. Section~\ref{DPforP}
provides a dynamic programming formulation and characterizes a notion
of optimal information utility. In Section~\ref{SecB} we provide three
lower bounds and two upper bounds on the optimal information utility.
The bounds are nonasymptotic and complementary for various values of
the parameters of the problem. Section~\ref{Optimality} states the
asymptotic consequence of the bounds obtained in Section~\ref{SecB}. In
particular, the obtained bounds are used to establish notions of
order and asymptotic optimality for the proposed policies (generalizing
that of~\cite{Chernoff59}); and
%(in Subsection~\ref{Gap});
characterize lower and upper bounds on the maximum achievable
information acquisition rate and the optimal reliability. %; and %(in
%Subsection~\ref{Rate}); and
%3) derive the advantage of causally and adaptively selecting sensing
%actions
%over the best open-loop (randomized) selection rule. %(in Subsection~
%Furthermore, the relative tightness of the bounds is used to
%establish a notion of asymptotic optimality for the proposed policies.
In Section~\ref{SecExmp} we investigate an important special case of
the active hypothesis testing, namely, the noisy dynamic search. In
Section~\ref{WeakenAsmp} we discuss the technical assumptions made in
our work and contrast them with the (weaker) assumptions in the
literature. More specifically, we show that our first technical
assumption weakens significantly one of the assumptions made in~\cite{Chernoff59}. On the other hand, our second technical assumption
is significantly stronger than the corresponding assumptions in the
literature. However, we show that while this assumption is critical in
obtaining the nonasymptotic lower and upper bounds of
Section~\ref{SecB}, it has no bearing on our asymptotic results in
Section~\ref{Optimality}.
%where the proposed heuristics are shown to be asymptotically optimal.
Finally, we conclude the paper and discuss future work in
Section~\ref{Discussion}. In the interest of brevity, we have chosen to
focus our analysis, provided in the \hyperref[AppLB]{Appendix}, on
Theorems~\ref{thmLB}--\ref{thmUB2}, whose results, to the best of our
knowledge, are entirely new and whose proofs require a substantially
different approach than those commonly available in the literature. In
contrast, the proofs of Propositions \ref{propLB1}--\ref{propUB2gen} as
well as Corollaries \ref{PeTo1}, \ref{CorLB2}, \ref{converse}--\ref{I2D2} follow similar lines of argument to the proofs
in the literature or in those obtained in the
\hyperref[AppLB]{Appendix} and are included in the form of a
supplemental article~\cite{Naghshvar-SuppA}.

%We close this section with a note on the notations used. In this paper,
\textit{Notation}: Let $[x]^+=\max\{x,0\}$. The indicator function
$\mathbf{1}_{\{A\}}$ takes the value~1 whenever event $A$ occurs, and 0
otherwise.
%A random variable is denoted by an upper case letter (e.g. $X$) and
%its realization is denoted by a lower case letter (e.g. $x$).
%Similarly, a random column vector and its realization are denoted by
%bold face symbols (e.g. $\mathbf{X}$ and $\mathbf{x}$).
%The $i$th element of vector $\mathbf{v}$ is denoted by $v_{i}$.
For any set $\mathcal{S}$, $\llvert  \mathcal{S} \rrvert $ denotes the
cardinality of $\mathcal{S}$.
%$\mathbf{1}_{j}$ denotes the vector of all zeros except for its
%$j^{th}$ component which takes value $1$.
%For a set~$\mathcal{A}$, let $\mathbb{P}(\mathcal{A})$ denote the
%collection of all probability distributions on
%elements of~$\mathcal{A}$, i.e., $\mathbb{P}(\mathcal{A}) = \{
%
All logarithms are in base~2. The entropy function on a vector
$\bolds{\rho}=[\rho_1,\rho
_2,\ldots,\rho_M] \in[0,1]^M$ %, $\sum_{i=1}^M \rho_i=1$,
is defined\vspace*{-1pt} as $H(\bolds{\rho})=\sum_{i=1}^{M} \rho_i \log
\frac{1}{\rho_i}$, % (1/{\rho_i})$,
%where $0 \log\frac{1}{0}$ is treated as 0.
with the convention that $0 \log\frac{1}{0} = 0$.
Finally, the Kullback--Leibler (KL) divergence between two probability
density functions $q(\cdot)$ and $q'(\cdot)$ on space $\mathcal{Z}$
%%denoted by $D(q\|q')$ where
is defined as $D(q\|q')=\int_{\mathcal{Z}} q(z) \log\frac {q(z)}{q'(z)}
\,dz$, with the convention $0 \log\frac{a}{0}=0$ and $b \log\frac
{b}{0}=\infty$ for $a,b\in[0,1]$ with $b\neq0$.
%Finally, let $\mathcal{N}(m,\sigma^2)$ denote a normal distribution
%with mean $m$ and variance $\sigma^2$.

%%%%----------------------------------------------------------------------------------------
%s2 #&#
\section{Problem setup and summary of the results} % and Dynamic
%Programming Formulation}
\label{PFDP} In Section~\ref{ProblemFormulation} we formulate the
problem of active sequential hypothesis testing, referred to as
Problem~\ref{probp} hereafter.
%Subsection~\ref{DPforP} provides the corresponding dynamic programming
%(DP) equation for Problem~(P)
%and characterizes an optimal policy for selecting actions.
Section~\ref{sum} states the main contributions of the paper and
provides a~summary of related works.

%s2.1 #&#
\subsection{Problem formulation}
\label{ProblemFormulation} Here, we provide a precise formulation of
our problem.
%for the following generalized active $M$-ary sequential hypothesis
%testing problem.

{\renewcommand{\theprob}{(P)}
%pr1 #&#
\begin{prob}[(Active sequential hypothesis testing)]\label{probp}%Active $M$-ary Hypothesis Testing
Let $\Omega_M=\{1,2,\break \ldots, M\}$. Let $H_i$, $i \in\Omega
    _M$, denote $M$ hypotheses of interest among which only one
    holds true. Let $\theta$ be the random variable that takes the
    value $\theta=i$ on the event that $H_i$ is true for $i
    \in\Omega_M$. We consider a Bayesian scenario with prior
    $\bolds{\rho}(0)=[\rho_1(0),\rho_2(0),\ldots,\rho_M(0)]$, that
    is, initially $P(\theta=i)=\rho_i(0)>0$ for all $i
    \in\Omega_M$. $\mathcal{A}_M$ is the set of all sensing actions
    which may depend on
$M$ and is assumed to be finite with $|\mathcal{A}_M| < \infty$. %$|
Let $\mathbb{P}(\mathcal{A}_M)$ denote the collection of all
probability distributions on elements of~$\mathcal{A}_M$, that is,
$\mathbb{P}(\mathcal{A}_M) = \{ \bolds{\lambda}
\in[0,1]^{|\mathcal{A}_M|}\dvtx  \sum_{a \in \mathcal{A}_M}
\lambda_a = 1 \}$. $\mathcal{Z}$~is the \textit{observation space}.
For all $a \in \mathcal{A}_M$, the observation kernel
$q^a_i(\cdot)$ (on~$\mathcal {Z}$) is the probability density
function for observation $Z$ when action~$a$ is taken and $H_i$ is
true.
%For any $a \in\mathcal{A}_M$, the \emph{one-step cost} of sensing
%action $a$ is given as $c^a$, $c^a \ge0$.
%We assume that the \emph{one-step cost} of all sensing actions is
%equal to 1.
We assume that observation kernels $\{q^a_i(\cdot)\}_{i\in\Omega
_M,a\in\mathcal{A}_M}$ are known and the observations are
conditionally independent over time.
Let ${L}$ denote the penalty (loss) for a wrong declaration, that
is, the penalty of selecting $H_j$, $j \neq i$, when $H_i$ is
true.\footnote{In general, we can define a loss matrix
$[L_{ij}]_{i,j\in\Omega_M}$, where $L_{ij}$ denotes the penalty
(loss) of selecting $H_j$ when $H_i$ is true.}
% Selecting the correct hypothesis is free of cost and hence,
%$l_{ii}=0$.
%The objective is to find the correct hypothesis at minimum cost.
Let $\tau$ be the stopping time at which the decision maker
retires. The objective is to find a sequence of sensing actions
$A(0)$, $A(1), \ldots, A({\tau-1})$,\footnote{We assume that $A(t)$
is selected as a (possibly randomized) function of\vspace*{-1pt}
$A^{t-1}_0:=[A(0),A(1),\ldots, A(t-1)]$ and
$Z^{t-1}_0:=[Z(0),Z(1),\ldots,Z(t-1)]$, that is, sensing actions
and observations up to time $t$.} a stopping time $\tau$ and a
declaration rule $d\dvtx \mathcal{A}_M^{\tau}
\times\mathcal{Z}^{\tau} \to\Omega_M$
that collectively minimize the expected total cost %:
%%%\ignore{$\mathbb{E}  [ \tau+ {L}{\mathbf{1}}_{\{d(A^{\tau
%%%},Z^{\tau}) \neq\theta\}}  ]$,
%%%% \begin{eqnarray}
%%%% \label{Objective}
%%%% \mathbb{E} \left[ \tau+ \L{\mathbf{1}}_{\{d(A^{\tau},Z^{\tau}) \neq
%%%%\theta\}} \right],
%%%% \end{eqnarray}
%%%where the expectation is taken with respect to the initial prior
%%%distribution $\bolds{\rho}(0)$ on $\theta$ as well as the
%%%distributions of action sequence, observation sequence, and the
%%%stopping time.
%%%%The total cost \eqref{Objective} can be written as
%%%The objective of Problem~(P) can be written as to }
%e1 #&#
\begin{equation}
\label{Objective2} % \mbox{ minimize }
\mathbb{E} [ \tau ]+ {L}\operatorname{Pe},
\end{equation}
where $\operatorname{Pe}=P(d(A^{\tau-1}_0,Z^{\tau-1}_0)
\neq\theta)$ denotes the probability of making a wrong declaration,
and the expectation is taken with respect to the initial prior
distribution $\bolds{\rho}(0)$ on $\theta$ as well as the
distributions of action sequence, observation sequence and the
stopping time.
%Based on \eqref{Objective2}, the region of interest for the parameters
%of the problem is
%$l > \log M$, since otherwise, the number of samples required to
%reduce the uncertainty about the hypotheses
%is higher than the penalty of random selection of the true hypothesis.}
\end{prob}}

%s2.2 #&#
\subsection{Overview of the results and summary of the related works}
\label{sum}

The first attempt to solve Problem~\ref{probp} goes back to Chernoff's
work on active binary composite hypothesis testing~\cite{Chernoff59}.
Chernoff proposed the following scheme to select actions: at each time
$t$, find the most likely true hypothesis, and then select an action
that can discriminate this hypothesis the best from each and every
element in the set corresponding to the alternative hypothesis. Much of the subsequent
literature extended this
approach~\cite{Bessler60,Albert61,Kiefer63,Blot73,Keener84,Lorden86,NitinawaratArxiv}.
\mbox{Chernoff} showed that as ${L}$~goes to infinity, the relative
difference between the expected total cost achieved by his proposed
scheme and the optimal expected total cost approaches zero, which he
termed as \textit{asymptotic
optimality}.\footnote{In~\cite{Chernoff59}, the\vspace*{-1pt} objective was to
minimize $c \mathbb{E}[\tau] + \operatorname{Pe}$ and the proposed
policy was shown to be asymptotically optimal as $c\to0$. It is
straightforward to show that for ${L}=\frac{1}{c}$, this problem
coincides with Problem~\ref{probp} defined in this paper. However, we
have chosen $\mathbb{E}[\tau]+{L}\operatorname{Pe}$ as an objective
function for Problem~\ref{probp} because of its interpretation as the
Lagrangian
relaxation of an information acquisition problem %of \cite{Burnashev76}
in which the objective is to minimize $\mathbb{E}[\tau]$ subject to
$\operatorname{Pe}\le\varepsilon$, where $\varepsilon>0$ denotes the
desired probability of error.}
One of the main drawbacks of Chernoff's asymptotic optimality notion
was his neglecting the complementary role of asymptotic analysis in
$M$. In particular, the notion of asymptotic optimality in ${L}$ falls
short in showing the tension between using an (asymptotically) large
number of samples to discriminate among a few hypotheses with
(asymptotically) high accuracy or an (asymptotically) large number of
hypotheses with a lower degree of accuracy. As a result, although the
scheme proposed in~\cite{Chernoff59}
and its subsequent extensions~\cite{Bessler60,Albert61,Kiefer63,Blot73,Keener84,Lorden86,NitinawaratArxiv} %NitinawaratICASSP12,
%NitinawaratArxiv}
are asymptotically optimal in~${L}$, their provable information
acquisition rate is restricted to zero.
%(potentially unbounded number of samples are used to acquire $\log M$
%bits of information).
Intuitively, the rate of information acquisition under any given
heuristic relates
to the ratio between $\log M$ and the expected number of samples: the
larger this ratio, the faster information is acquired.

As elaborated in Section~\ref{Rate}, to obtain asymptotic
characterization of the optimal expected total cost %stopping time $
in a nonzero rate regime, it is important to propose schemes which
scale optimally with $M$ as well.
In his seminal paper~\cite{Burnashev76}, Burnashev tackled the primal
(constrained) version of Problem~\ref{probp} in the context of channel
coding with feedback,
%(in Section~\ref{VLcoding} we explain why channel coding with feedback
%can be interpreted as a special case of Problem~(P))
and provided lower and upper bounds on the expected number of samples
(or, equivalently, channel uses) required
to convey one of $M$ uniformly distributed messages over a discrete
memoryless channel (DMC) with a desired probability of error. % less
%than some $\varepsilon> 0$.
The lower bound identified the dominating terms in both number of
messages and error probability, hence characterized the optimal
reliability function (also known as the error exponent) in addition to
the feedback capacity (which was known to coincide with the Shannon
capacity~\cite{Shannon56}).
In this paper, we generalize\footnote{In~\cite{Burnashev80}, Burnashev
attempted to tackle the problem of active sequential hypothesis testing
by Chernoff~\cite{Chernoff59}. However, the sensing actions
in~\cite{Burnashev80} were allowed to be functions of the true
hypothesis,~$\theta$, which, in general, is not observable in the
active testing setting~\cite{Chernoff59}.
%with the exception of the problem of variable-length coding with
%feedback as we discuss in Section~\ref{VLcoding}.
In this sense, \cite{Burnashev80} only extends Burnashev's earlier
work~\cite{Burnashev76} on variable-length coding over a discrete
memoryless channel (DMC) with feedback to allow for more general
channels.} this lower bound to the problem of active sequential
hypothesis testing, that is, Problem~\ref{probp}:
%In particular, we characterize the solution to Problem~(P) in terms of
%$\L$ and $M$.
%The first contribution of this result is to generalize Burnashev's
%analysis beyond
%variable-length coding context to the general setup of Chernoff:
%
\begin{itemize}
\item We derive three lower bounds on the expected total
    cost~(\ref{Objective2}).
%and state ? bounds obtained in~\cite{Chernoff59}.
The bounds hold for all prior beliefs and are nonasymptotic and
complementary for various values of ${L}$ and $M$. In
Section~\ref{Optimality} these bounds are collectively used to
generalize the (information theoretic) notions of achievable
communication rate~\cite{CoverBook2nd} and error
exponent~\cite{Gallager68} to the context of active sequential
hypothesis testing.
% a notion of information acquisition rate and reliability functions.

\item The first and second lower bounds identify the dominating
    terms in $L$ and hence are useful in establishing asymptotic
    optimality of order-1 (due to Chernoff~\cite{Chernoff59}) and
    order-2 in $L$. Furthermore, from an information theoretic
    viewpoint, these bounds are used to characterize an upper bound
    on the reliability function (error exponent) at zero rate.

\item The third lower bound characterizes the dominating terms of
    growth in the optimal expected total cost in terms of ${L}$ and
    $M$ simultaneously. We use this as a converse (in a fashion
    somewhat similar to Shannon's channel coding
    converse~\cite{CoverBook2nd}) to derive an upper bound
    $\widebar{I}_{\max}$ on the maximum achievable \mbox{information}
    acquisition rate. Additionally, this lower bound allows us to
    provide an upper bound on the reliability function (error
    exponent) for all rates $R\in[0,\widebar{I}_{\max}]$, and
    establish order optimality in $M$ as a necessary condition
    for any policy which achieves nonzero information acquisition
    rate.
\end{itemize}

In addition to a lower bound on an expected number of samples,
Burnashev proposed a coding scheme with two phases of operation
whose performance provides a tight upper bound (in both number of
messages and error probability).
%(in both $\L$ and $M$).
It is interesting to note that the scheme of Chernoff, if specialized
to channel coding with feedback,
coincides with the second phase of Burnashev's scheme and is of a
repetition code nature.
This means that while the first phase of Burnashev's scheme can achieve
any information rate up to the capacity of the channel,
Chernoff's one-phase scheme has a rate of information acquisition equal
to zero.
Inspired by Burnashev's coding scheme, we also obtain
%two upper bounds on the expected total cost \eqref{Objective2}. These
%bounds are derived via asymptotic analysis of
%$\tilde{\pi}_2$:
two heuristic two-phase policies $\tilde{\pi}_1$ and $\tilde{\pi }_2$
whose nonasymptotic analysis in Proposition~\ref{propUB1} and
Theorem~\ref{thmUB2} provides two upper bounds on the optimal
performance:
\begin{itemize}
% \item Policy $\tilde{\pi}_1$, in its first phase, selects actions in
%a way that
%all pairs of hypotheses can be distinguished from each other;
%while its second phase coincides with Chernoff's scheme~
%%(and also the schemes proposed in \cite{Srivastava11CDC,%NitinawaratICASSP12,AtiaISIT12,NitinawaratArxiv})
%where only the pairs including the most likely hypothesis are
%considered.
%The second phase of $\tilde{\pi}_1$ is shown to ensure its asymptotic
%optimality in $\L$;
%while its first phase in a very natural manner relaxes the technical
%assumption in~\cite{Chernoff59}
%where all actions are assumed to discriminate between all hypotheses
%pairs or the need for the infinitely often reliance on randomized
%action deployed in~\cite{NitinawaratArxiv} in order to ensure
%sufficient discrimination among hypotheses.
\item Policy $\tilde{\pi}_1$ is a simple two-phase modification of
    Chernoff's scheme in which testing for the maximum likely
    hypothesis is delayed and contingent on obtaining a certain
    level of confidence. More specifically, in its first phase,
    $\tilde{\pi}_1$ selects actions in a way that all pairs of
    hypotheses can be distinguished from each other, while its
    second phase coincides with Chernoff's scheme~\cite{Chernoff59}
    where only the pairs including the most likely hypothesis are
    considered. The second phase of $\tilde{\pi}_1$ ensures its
    asymptotic optimality in ${L}$, while its first phase in a very
    natural manner weakens the technical assumption
    in~\cite{Chernoff59} in which all actions are assumed to
    discriminate between all hypotheses pairs or the need for the
    infinitely often reliance on suboptimal randomized action
    deployed in~\cite{Chernoff59,NitinawaratArxiv}.
%
%benchmark for comparison when considering Chernoff's asymptotic
%regime.} \tj{this is not necessary now i think}
%to enable our identifying the adaptivity gain in Section~

\item Policy $\tilde{\pi}_2$ is only shown to be asymptotically
    optimal in $L$ under a stronger condition, which is later shown
    to be satisfied in the important cases of binary hypothesis
    testing and noisy dynamic search in Section~\ref{SecExmp},
    however, with the advantage that here for a fixed $M$ the
    asymptotic optimality \cite{Chernoff59} can be strengthened to
    a higher order. In particular, in Section~\ref{GapL}, we show
    that when $\tilde{\pi }_2$ is asymptotically optimal it
    achieves a bounded difference with the optimal performance.
% binary hypothesis testing, noisy dynamic search, and variable-length
%coding with feedback.
Furthermore, under a technical condition, policy $\tilde{\pi}_2$
can ensure that information acquisition occurs at a nonzero rate.
Mathematically, this means that, under policy $\tilde{\pi}_2$, the
expected total cost~(\ref{Objective2}) grows in ${L}$ and $M$ in an
order optimal fashion establishing a lower bound on the maximum
achievable information acquisition rate $\underline{I}_2
\le\widebar{I}_{\max}$ as well as a lower bound on the optimal
reliability function (optimal error exponent) for all rates
$R\in[0,\underline{I}_2]$.
\end{itemize}

%The above results are also used to answer a fundamental question
%regarding the significance of adaptive decision
%making. In particular, specializing the obtained lower bounds to the
%open-loop setup along with our policy $\tilde{\pi}_1$,
%we investigate the benefit of adapting sensing actions at any decision
%epoch. We show that almost in all practical settings
%the adaptivity gain grows logarithmically with penalty $L$. Such a
%characterization complements the more recent
%work on multi-stage policies (introduced in~
%take a retire/declare action only at the end of each stage (of
%potentially unequal length).
%Extending our characterization of the adaptivity gain to quantifying
%the loss in performance
%introduced by the multi-stage decision making constraint remains an
%area of future work.

%discussed in Section~\ref{secNDS} is a problem of independent and
%extensive interest where Chernoff-type schemes
%optimize the reliability (error exponent) at zero rate; while
%generalizations of binary search have been proposed to address the
%speedy identification of the true hypothesis
%solutions here rely on info max or something like that at all? this is
%a bit too thin; I think can be beefed up!)}
%We treat this example in Section~\ref{secNDS} extensively to
%illustrate contributions of our work as well as highlight the
%rate--reliability trade-off.}

To illustrate contributions of our work as well as highlight the
rate--reliability trade-off, we treat the problem of noisy dynamic
search in Section~\ref{secNDS}. This problem is of independent and
extensive interest, and arises in a variety of fields from fault
detection to whereabouts search to noisy group testing.
We specialize the results obtained in the earlier sections for the
general active hypothesis testing, and discuss our findings in the
context of other solutions in the literature.
Particularly, in the case of size-independent Bernoulli noise, the
upper bound corresponding to policy $\tilde{\pi}_2$ is shown to be
asymptotically tight in both ${L}$ and $M$, hence ensuring the maximum
acquisition rate and reliability simultaneously, but there is no
guarantee on the tightness of the bounds for general noise models. The
potentially growing gap between the lower and upper bounds obtained
here, in particular, underline the significant complications of
acquiring information in the general active hypothesis testing over
that of (variable-length coding with feedback)~\cite{Burnashev80}. For
instance, while in the channel coding context the maximum information
rate and reliability are fully known and match that of channel capacity
and error exponent, they remain largely uncharacterized, beyond our
bounds here, even in the practically relevant problem of a noisy
dynamic search.

As briefly discussed in the \hyperref[sec1]{Introduction}, the above results have all
been obtained under an important technical assumption which is stronger
than those commonly made in the literature. However, we will show that
this assumption can be significantly weakened. More precisely, we show
that our original technical assumption can be replaced with one that is
weaker, to the best of our knowledge, than all other assumptions in the
literature
\cite{Bessler60,Albert61,Kiefer63,Blot73,Lorden77,Burnashev76,NitinawaratArxiv},
and, in particular, subsumes that of \cite{Chernoff59}, to obtain a set
of (nonasymptotic) bounds which are looser than those obtained in
Section~\ref{SecB}. On the other hand, these looser (nonasymptotic)
bounds are shown to have similar dominating terms to those obtained in
Section~\ref{SecB}, and hence ensure the validity of our asymptotic
results in Section~\ref{Optimality}.

%----------------------------------------------------------------
%s3 #&#
\section{Dynamic programming and characterization of an optimal policy}
\label{DPforP}

In this section we first derive the corresponding dynamic programming
(DP) equation for Problem~\ref{probp}.
%Then we provide some numerical examples to gain intuition about
%characteristics of the optimal solution.
%Finally we use these examples to present an overview of our
%contributions.
From the DP solution, we characterize an optimal policy for
Problem~\ref{probp}.
%we present an overview of the results in \cite{ISIT10} regarding the
%DP solution and also the sufficient condition for reduction to passive
%testing.

The problem of active $M$-ary hypothesis testing is a partially
observable Markov decision problem (POMDP) where the state is static
and observations are noisy. It~is known that any POMDP is equivalent to
an MDP with a compact yet uncountable state space, for which the belief
of the decision maker about the underlying state becomes an information
state~\cite{Kumar}.
%Specifically, let $\Theta=\sigma\{ \theta_1, \theta_2, \ldots,
%Specifically, let $\Theta$ be the $\sigma$-algebra of the events $\{
In our setup, thus, the information state at time $t$ is the belief
vector $\bolds{\rho}(t)$ whose $i$th element is the conditional
probability of hypothesis $H_i$ to be true given the initial belief and
all the observations and actions up to time $t$, that is,
$\rho_i(t):=P(\{\theta=i\}|A^{t-1}_0,Z^{t-1}_0)$. Accordingly, the
information state space is defined as $\mathbb {P}(\Omega_M):=
\{\bolds{\rho} \in[0,1]^M\dvtx  \sum_{i=1}^{M} {\rho}_i = 1  \}$ and
the optimal expected total cost can be defined as follows.

\begin{definition}
For all $\bolds{\rho} \in\mathbb{P}(\Omega_M)$, let functional
$V^*(\bolds{\rho})$, hereafter referred to as the \textit{optimal value
function}, denote the optimal expected total cost~(\ref{Objective2}) of
Problem~\ref{probp} given the Bayesian prior $\bolds{\rho}$. In other
words, $V^*(\bolds{\rho}):=\min \{\mathbb{E} [ \tau ]+
{L}\operatorname{Pe} \}$ given the initial belief $\bolds{\rho}$, where
the minimization is taken over the stopping time~$\tau$, the sequence
of actions and observations, and the declaration rule.
\end{definition}

A general approach to solving Problem~\ref{probp} is to provide a
functional characterization of $V^*$: given $V^*$ in its functional
form, the optimal expected total cost for Problem~\ref{probp} can be
obtained by a simple evaluation of $V^*$ at the initial belief
$\bolds{\rho}(0)$. Next we state a dynamic programming equation which
characterizes $V^*$.

% where $\Theta$ is the \mbox{$\sigma$-algebra} generated by random
%variable $\theta$.
%Due to the i.i.d. (over time) nature of the observation process, the
%future belief vector is only a function of the current belief vector $
To obtain the dynamic programming equation, consider a single step of
the problem. In one sensing step, the evolution of the belief vector
follows Bayes' rule and is given by ${\bolds{\Phi}}^a$, a measurable
function from $\mathbb {P}(\Omega_M) \times\mathcal{Z}$ to
$\mathbb{P}(\Omega_M)$ for all $a \in\mathcal{A}_M$:
%e2 #&#
\begin{equation}
\label{PhiDef} %{\bolds{\Phi}}^a(\rho,z)(\theta_i)=\rho(\theta_i) \frac{q^a(z|
{\bolds{\Phi}}^a(\bolds{\rho},z):= \biggl[ \rho_1
\frac
{q^a_1(z)}{q^a_{\bolds{\rho}}(z)}, \rho_2 \frac
{q^a_2(z)}{q^a_{\bolds{\rho}}(z)}, \ldots,
\rho_{M} \frac
{q^a_{M}(z)}{q^a_{\bolds{\rho}}(z)} \biggr],
\end{equation}
where $q^a_{\bolds{\rho}}(z)=\sum_{i=1}^{M} \rho_i q^a_i(z)$, and
${\bolds{\Phi}}^a(\bolds{\rho},z)=\bolds{\rho}$ if $q^a_{\bolds{\rho}}(z)=0$. %$0 < q^a_{\bolds{
In other words, if $\bolds{\rho} \in\mathbb{P}(\Omega_M)$ is an a
priori distribution, ${\bolds{\Phi}}^a(\bolds{\rho},z)$ gives us the
posteriori distribution when sensing action $a$ has
been taken and $z$ has been observed. %Note that the posterior
%distribution is strongly dependent on the sensing action~$a$.

We define a Markov operator $\mathbb{T}^a$, $a \in\mathcal{A}_M$, such
that for any measurable function $g\dvtx  \mathbb{P}(\Omega_M)
\to\mathbb{R}$,
%e3 #&#
\begin{equation}
\label{OperatorT} \bigl(\mathbb{T}^a g\bigr) (\bolds{\rho}):= \int
g\bigl( {\bolds{\Phi }}^a(\bolds{\rho},z) \bigr)
q^a_{\bolds{\rho}}(z) \,dz.
\end{equation}
%
%Given that $\bolds{\rho}$ is an apriori belief and action $a$ has
%been taken,
%$(\mathbb{T}^a g)(\bolds{\rho})$ is the expected value of
%function $g$ on the space of posterior beliefs,
%where the computation of the posterior belief follows Bayes{\chr"92}
%rule as shown in~(\ref{PhiDef}).
%%
%Note that using operator $\mathbb{T}^a$, one can compute the mutual
%information between $\theta$ with distribution $\bolds{\rho}$
%and observation $Z$ under action $a$ with distribution $q^a_{
%I(\bolds{\rho};q^a_{\bolds{\rho}}) &:= H(\bolds{\rho})
%- \int H({\bolds{\Phi}}^a(\bolds{\rho},z)) q^a_{\bolds{
%&= H(\bolds{\rho}) - (\mathbb{T}^a H)(\bolds{\rho}).

%For all $\bolds{\rho} \in\mathbb{P}(\Omega_M)$, let functional
%$V^*(\bolds{\rho})$ denote the optimal expected total cost~
Note that at any given information state $\bolds{\rho}$, taking sensing
action $a\in\mathcal{A}_M$ followed by the optimal policy results in
expected total cost $1 + (\mathbb{T}^{a} V^{*})(\bolds{\rho})$, where
$1$ denotes the one unit of time spent to take the sensing action and
collect the corresponding observation sample, and $(\mathbb{T}^{a}
V^{*})(\bolds{\rho})$ is the expected value of $V^*$ on the space of
posterior beliefs; while declaration $j$ results in expected cost
$(1-\rho_j) L$ where $(1-\rho_j)$ is the probability that hypothesis
$H_j$ is not true, and $L$ is the penalty of making a wrong
declaration.
%results in expected total cost $1 + (\mathbb{T}^{a} V^{*})(\bolds{
%while declaration $j$ results in penalty $(1-\rho_j) \L$.
This intuition, while relying on the compactness of $\mathbb{P}(\Omega
_M)$ to treat various measurability issues, can be formalized in the
following dynamic programming equation.

%
%fa1 #&#
\begin{fact}[(Proposition 9.8 in~\cite{Bertsekas07})]
%Let functional $V^*:\mathbb{P}(\Omega_M) \to\mathbb{R}_+$ be the
%minimal
The optimal value function $V^*$ % $V^*\dvtx  \mathbb{P}(\Omega_M) \to
satisfies the following fixed point equation:
%is a solution to the following fixed point equation:
%a functional solving the following fixed point equation:
%e4 #&#
\begin{eqnarray}
\label{ValueF} V^{*}(\bolds{\rho}) &=& \min \Bigl\{ 1 + \min
_{a \in\mathcal
{A}_M} \bigl(\mathbb{T}^{a} V^{*}\bigr) (
\bolds{\rho}), \min_{j\in\Omega
_M} (1-\rho_j) {L} \Bigr
\}.
\end{eqnarray}
%
%Then $V^{*}(\bolds{\rho}(0))$, referred to as the optimal value
%function, is equal to the minimum expected total cost in Problem~(P)
%with the prior belief~$\bolds{\rho}(0)$.
\end{fact}

%in a Markovian, stationary, and deterministic fashion without loss of
%optimality \cite{Strauch66}, Theorem 9.1,.}
%is a mapping from the information state space $\mathbb{P}(\Omega_M)$
%to action space
%$\mathcal{A}_M$ based on which sensing actions $A(t)$, $t=0,1,\ldots,
%stopping time $\tau$, and declaration rule $d$ are selected.
%From \cite{Strauch66}, there exists a Markov stationary deterministic
%policy
%which minimizes (\ref{Objective}) and is referred to as an

\begin{definition}
A Markov stationary \textit{policy} is a stochastic kernel from the
information state space $\mathbb{P}(\Omega_M)$ to $\mathcal{A}_M
\cup\{d\}$ describing the conditional distribution on sensing actions
$A(t)$, $t=0,1,\ldots,\tau-1$ and stopping time $\tau$ (the choice of
declaration $d$ marks the stopping time $\tau$). In other words, under
policy $\pi$, the probability that action $a$ is selected at belief
state $\bolds{\rho}$ is given by $\pi(a|\bolds{\rho})$.
%In other words, under policy $\pi$, action $a$ is selected at belief
%state $\bolds{\rho}$
%with probability $\pi(a|\bolds{\rho})$.
\end{definition}

%A policy $\pi$ is referred to as Markov stationary
%there exists an action $a \in\mathcal{A}_M \cup\{d\}$ for which $
%%From \cite{Strauch66}, there exists a Markov stationary deterministic
%policy
%%which minimizes (\ref{Objective}) and is referred to as an

%Next we define a class of policies which do not fully utilize the
%observation outcomes.
%%A policy $\pi$ is referred to as open-loop or nonadaptive if $\pi(
%A policy $\pi$ is referred to as \emph{open-loop} or
%independent
%of the observation outcomes (hence independent of the belief state).
%For a given vector $\bolds{\lambda} \in\mathbb{P}(
%selects sensing actions $a \in\mathcal{A}_M$ with probability $\pi_{
%independent of $\bolds{\rho}$ until the stopping time $\tau$ is
%reached.

%From \cite{Strauch66}, there exists a Markov stationary deterministic
%policy
%From Theorem~14 in \cite{Bertsekas79}, there exists a Markov
%stationary deterministic policy
As shown in Corollary~9.12.1 in~\cite{Bertsekas07},
equation~(\ref{ValueF}) provides a characterization of an optimal
Markov stationary deterministic policy $\pi^*$ for Problem~\ref{probp}
as follows: sensing action $a^*=\argmin_{a
\in\mathcal{A}_M}(\mathbb{T}^{a} V^{*})(\bolds{\rho})$ is the least
costly sensing \mbox{action}, resulting in $1 + \min_{a \in \mathcal{A}_M}
(\mathbb{T}^{a} V^{*})(\bolds{\rho})$, hence is the optimal action to
take unless wrongly declaring $H_{i^*}$, where
$i^*=\argmin_{j\in\Omega_M} (1-\rho_j) {L}$, is even less costly, in
which case it is optimal to retire and declare $H_{i^*}$ as the true
hypothesis.

%
%re1 #&#
\begin{remarks}
\label{IntrstReg} It follows from~(\ref{ValueF}) that if
$\min_{j\in\Omega_M}(1-\rho _j) {L}\le1$, then we have a full
characterization of $V^{*}(\bolds{\rho})=\min_{j\in\Omega_M}(1-\rho_j)
{L}$ and the optimal policy.
%the further reduction of the probability of error is not worth taking
%one more sensing action.
%and it is optimal to retire.
Therefore, the region of interest in our analysis is restricted to
${L}> 1$ and $\mathbb{P}_{{L}}(\Omega_M):=\{\bolds{\rho}\in\mathbb
{P}(\Omega_M)\dvtx  \min_{j\in\Omega_M} (1-\rho_j) {L}> 1\}$.
%Therefore, the region of interest for the prior belief is
%$\{\bolds{\rho}\in\mathbb{P}(\Omega_M)\dvtx  \min_{j\in\Omega_M} (1-
\end{remarks}

Before we close this section, we provide the following lemma.

%Suppose there exists a functional $V\dvtx \mathbb{P}(\Omega_M) \to
%$$V(\bolds{\rho}) \le\min\{ 1+ \min_{a \in
%Then $V(\bolds{\rho}) \le~V^{*}(\bolds{\rho})$ for all $
%where $V^*$ is a solution to~\eqref{ValueF}.

%
%le1 #&#
\begin{lemma}
\label{lemVLB} Suppose there exist $\beta> 0$ and a functional $V\dvtx
\mathbb{P}(\Omega _M) \to\mathbb{R}_+$ such that for all belief vectors
$\bolds{\rho} \in\mathbb{P}(\Omega_M)$,
\[
V(\bolds{\rho}) \le\min\Bigl\{ \beta+ \min_{a \in
\mathcal{A}_M} \bigl(
\mathbb{T}^{a} V\bigr) (\bolds{\rho}), \min_{j\in\Omega_M} (1-
\rho_j)\beta{L}\Bigr\}.
\]
Then $V^{*}(\bolds{\rho})\ge\frac{1}{\beta}V(\bolds{\rho})$ for all
$\bolds{\rho} \in\mathbb{P}(\Omega_M)$.
%where $V^*$ is a solution to~\eqref{ValueF}.
\end{lemma}

The proof is provided in the supplemental article~\cite{Naghshvar-SuppA}, Section~1. %\ref{supp:VLB}. %in Appendix~\ref{AppPuterman}.
%{\textcolor{green}{I think we can prove this without using assumption
%1}}
%The proof follows similar lines as the proof of Proposition~7.3.4 of

%%%%%%%%%%%%%%%%%%%%%%%%%%%%%%%%%%%%%%%%%%%%%%%%%%%%%%%%

%%%%---------------------------------------------------------------------------------------------
%Results
%s4 #&#
\section{Performance bounds}\label{SecB}

%In this section we will consider the problem of active binary
%hypothesis testing ($M=2$).
%Furthermore, we assume that all actions have the same one-step cost,
%i.e., $c^a=c$ for all $a \in\mathcal{A}_M$.

%As discussed earlier, finding an optimal policy $\pi^*$ for
%Problem~(P) requires knowledge of the optimal value function $V^*$.
%In lieu of numerical approximation of~(\ref{ValueF}) using value
%iteration techniques~\cite{Puterman},
%or deriving a closed-form for $V^*$, in Subsections~\ref{SecLB} and~
%we use Lemma~\ref{lemVLB} and heuristic policies
%to find lower and upper bounds for the value function $V^{*}$
%respectively.
In lieu of numerical approximation of or derivation of a closed form
for $V^*$, in Section~\ref{SecLB} we use Lemma~\ref{lemVLB} to find
lower bounds for the value function $V^{*}$. In Section~\ref{SecUB} we
analyze two heuristic schemes to achieve upper bounds for $V^{*}$.
%First of all, these lower and upper bounds characterize $V^*$ in an
%order sense.
%Secondly, we use the bounds in Subsection~\ref{Optimality} to quantify
%the advantage of adaptive decision making
%over the best (randomized) open-loop solution by lower bounding the
%adaptivity gain.
%In particular, it is shown that in most practical settings, adaptive
%decision making
%provides a logarithmic reduction in cost as the penalty of wrong
%declaration grows
%(or equivalently, as the probability of erroneous detection approaches
%zero).

%Before we proceed, we need the following definition. %of the
%Kullback--Leibler divergence between two probability density functions.
%
%variable $\theta$ and observation $Z$ when sensing action $a$ is taken
%and the belief state is $\bolds{\rho}$ is denoted by
%% $I(\bolds{\rho};q^a_{\bolds{\rho}})$
%I(\bolds{\rho};q^a_{\bolds{\rho}}) %&= H(\bolds{\rho})
%- H({\bolds{\Phi}}^a(\bolds{\rho},Z))\\
%: = H(\bolds{\rho}) - \int H({\bolds{\Phi}}^a(\bolds{
%where $H(\bolds{\rho})=\sum_{i=1}^{M} \rho_i \log(1/{\rho_i})$
%is the entropy function.

%The \emph{Kullback--Leibler (KL) divergence} between two probability
%density functions $q(\cdot)$ and $q'(\cdot)$ is denoted by $D(q\|q')$
%where
%$$D(q\|q')=\int q(z) \log\frac{q(z)}{q'(z)} \,dz.$$
%
%The \emph{Jensen-Shannon (JS) divergence} among probability density
%functions $q_1,q_2,\ldots,q_M$ with respect to ${\bolds{\rho}=\{
%is defined as $$JS(\bolds{\rho},q_1,q_2,\ldots,q_M)=\sum_{i=1}^M
%where $q_{\bolds{\rho}}(z)=\sum_{i=1}^M \rho_i q_i(z)$.
%
%It is known that $I(\bolds{\rho};q^a_{\bolds{\rho}})=JS(

We have the following technical assumptions:

%
%as1 #&#
\begin{assumption}
\label{KL0}
%For any hypothesis $i$, there exists an action $a$, $a \in
For any two hypotheses $i,j\in\Omega_M$, $i \neq j$, there exists an
action $a$, $a \in\mathcal{A}_M$, such that $D(q^a_i \| q^a_j) > 0$.
\end{assumption}

%
%as2 #&#
\begin{assumption}
\label{Jump} There exists $\xi_M < \infty$ such that
\[
\max_{i,j\in\Omega_M}\ \max_{a \in\mathcal{A}_M}\ \sup_{z \in\mathcal{Z}} \log\frac{q^a_i(z)}{q^a_j(z)} \le
\xi_M.
\]
%
% < \infty$.
\end{assumption}
%
%There exists a $\gamma> 0$ such that
%$l \ge\frac{\log M}{\max_{a\in\mathcal{A}_M} \max_{
%I(\hat{\bolds{\rho}};q^{a}_{\hat{\bolds{\rho}}}) - \gamma}$,
%%
%where $I(\bolds{\rho};q^a_{\bolds{\rho}})= H(\bolds{
%and observation $Z$ under action $a$.
%

Assumption~\ref{KL0} ensures the possibility of discrimination between
any two hypotheses, hence ensuring Problem~\ref{probp} has a meaningful
solution. Assumption~\ref{Jump} implies that no two hypotheses are
fully distinguishable using a single observation sample.
Assumption~\ref{Jump} is a technical one which enables our
nonasymptotic characterizations, however, in Section~\ref{WeakenAsmp}
we discuss the consequence of weakening this assumption in detail.
%the proof of the upper bounds.
%However, at the cost of increasing notation and more complicated
%analysis,
%Assumption~\ref{Jump} can be replaced by a more general assumption as
%those considered in~\cite{Burnashev80}, p.~421,.
%%%\ignore{ We believe that, at the cost of increasing notation, standard
%%%techniques as those provided in~\cite{Chernoff59,Burnashev80,Bentkus08}
%%%can be applied to generalize the bounds when Assumption~\ref{Jump} does
%%%not hold.}

%to have a %solution with information acquisition rate strictly above
%zero.}
%non-trivial solution as both $\L$ and $M$ increase.
%Next lemma precisely illustrates the necessity of Assumption~
%in our asymptotic analysis as $\L$ and $M$ tend to infinity.
%More discussion on this is provided in Section~\ref{Rate}.}

%Let $\frac{\log M}{l} > \max_{a\in\mathcal{A}_M} \max
%_{\hat{\bolds{\rho}}\in\mathbb{P}_{\L}(\Omega_M)}
%I(\hat{\bolds{\rho}};q^{a}_{\hat{\bolds{\rho}}})$,
%and suppose the decision maker has a uniform prior belief about the
%hypotheses.
%As $M \to\infty$, the optimal policy randomly guesses the true
%hypothesis without collecting any observation sample
%and hence, $\Pe$, the probability of making a wrong declaration,
%approaches 1.
%
%The proof of Lemma~\ref{PeTo1} is provided in Appendix~
%}

%--LOWER
%%%BOUND--------------------------------------------------------------------------
%s4.1 #&#
\subsection{Lower bounds for $V^{*}$}\label{SecLB}

%Under Assumption~\ref{KL0} and for $\L> 1$, $V^*(\bolds{\rho})
%{\max_{a\in\mathcal{A}_M} D(q^{a}_i\|q^{a}_j)} - K'_1
%%
%and $K'_1$ is a constant independent of $\L$.

%
%th1 #&#
\begin{theorem}
\label{thmLB} Under Assumption~\ref{KL0} and for ${L}> 1$ and
$\bolds{\rho} \in\mathbb{P}_{{L}}(\Omega_M)$,
\[
V^*(\bolds{\rho}) \ge\underline{V_1}(\bolds{\rho}):=
\Biggl[ \sum_{i=1}^{M} \rho_i
\max_{j \neq i} \frac{\log
((1-L^{-1})/L^{-1}) - \log (\rho_i/\rho_j)} {
\max_{a\in\mathcal{A}_M} D(q^{a}_i\|q^{a}_j)} - K'_1
\Biggr]^+,
\]
where $K'_1$ is a constant independent of ${L}$ whose closed form is
given in the supplemental article~\cite{Naghshvar-SuppA},
equation~(144).
%whose closed-form is given in \eqref{K1close} in the supplemental
%article~\cite{Naghshvar-SuppA}, Section~\ref{supp:Claims},.
\end{theorem}

The proof of Theorem~\ref{thmLB} is provided in
\hyperref[AppLB]{Appendix}.
%The proof of Proposition~\ref{thmLB} is based on Lemma~\ref{lemVLB}
%and is provided in Appendix~\ref{AppLB1}.

Following Chernoff's approach (Theorem~2 in \cite{Chernoff59}),  and for
large values of ${L}$, the lower bound can be tightened as follows:
%(in an asymptotic sense) as:
%
%pr1 #&#
\begin{proposition}\label{propLB1} %[Proposition~3 in~\cite{CISS2012}]
Under Assumptions~\ref{KL0} and~\ref{Jump}, and for ${L}> 1$,
$\bolds{\rho} \in\mathbb{P}_{{L}}(\Omega_M)$, and arbitrary
$\delta\in(0,1)$,
%Under Assumptions~\ref{KL0} and~\ref{Jump} and for $\L> 1$ \tj{and $
%V^*(\bolds{\rho})&\ge
%{\max_{\bolds{\lambda}\in\mathbb{P}(\mathcal{A}_M)} \min
%_{j\neq i} \sum_{a\in\mathcal{A}_M} \lambda_{a}
%D(q^{a}_i\|q^{a}_j)} \right. \\ %- o(M \log\L) \right]^+,
%& \hspace*{1.85in} \left. - \frac{o(M \log\L)}{\big( \max_{
%where $\frac{o(M \log\L)}{M \log\L} \to0$ as $M \L\to\infty$.
%e5 #&#
\begin{eqnarray*}
V^*(\bolds{\rho})
&\ge& \Biggl[ \sum_{i=1}^{M}
\rho_i \frac{ [(1-\delta)\log (L/(K'\log2L)) - \max_{j \neq i} \log(\rho_i/\rho
_j) ]^+} {\max_{\bolds{\lambda}\in\mathbb{P}(\mathcal{A}_M)}
\min_{j\neq i} \sum_{a\in\mathcal{A}_M} \lambda
_{a} D(q^{a}_i\|q^{a}_j) + \delta}
\\
&&\hspace*{92pt}{}\times \biggl(1-\frac{2M (K'\log
2{L}/L)^{\delta}}{\rho_i} \biggr) -
\frac{M \xi_M^2}{\delta
^2} \Biggr]^+,
\end{eqnarray*}
where $K'$ is a constant independent of $\delta$ and ${L}$ whose closed
form is given in the supplemental article~\cite{Naghshvar-SuppA},
equation~(81).
\end{proposition}
%
%Fact~\ref{LB3} complements the results of \cite{Chernoff59,%NitinawaratArxiv} in a total cost (and Bayesian) sense
%and takes into account the asymptotic in $M$ as well.
%The proof of Fact~\ref{LB3} follows closely the proof of Theorem~2 in

The proof of Proposition \ref{propLB1} is provided in the supplemental
article \cite{Naghshvar-SuppA}, Section~5.1.

Next we provide another lower bound which is more appropriate for large
values of $M$.
%Before we proceed, we need the following notations.
Let $I(\bolds{\rho};q^a_{\bolds{\rho}})= H(\bolds{\rho}) -
(\mathbb{T}^a H)(\bolds{\rho})$ denote the mutual information between
$\theta\sim\bolds{\rho}$ and observation $Z$ under action $a$.
Let $D_{\max}(M):=\max_{i,j\in\Omega_M} \max _{a \in\mathcal{A}_M}
D(q_i^{a}\|q_j^{a})$, $I_{\max}(M):=\max_{a \in\mathcal{A}_M} \max
_{\hat{\bolds{\rho}}\in\mathbb{P}(\Omega_M)}
I(\hat{\bolds{\rho}};q^{a}_{\hat{\bolds{\rho}}})$,\break and
$\alpha({L}, M):=\frac{M-1}{M-1+2^{{L}I_{\max}(M)}}$.

%
%th2 #&#
\begin{theorem}
\label{thmLB2} Under Assumption~\ref{KL0} and for ${L}> 1$ and
$\bolds{\rho} \in\mathbb{P}_{{L}}(\Omega_M)$,
%Assumptions~\ref{KL0} and~\ref{Jump}
\begin{eqnarray*}
V^*(\bolds{\rho})
&\ge& \biggl[\frac{H(\bolds{\rho})-H([\alpha({L},M),1-\alpha
({L},M)])-\alpha({L},M)\log(M-1)}{I_{\max}(M)}
\\[-4pt]
&&\hspace*{218pt}{}+\alpha ({L},M){L} \biggr]^+.
\end{eqnarray*}
Furthermore, under Assumptions~\ref{KL0} and~\ref{Jump}, and for ${L}>
\max\{1,\frac{\log M}{I_{\max}(M)}\}$ and arbitrary $\delta\in(0,0.5]$,
%
%e6 #&#
\begin{eqnarray*}
V^*(\bolds{\rho}) &\ge&\underline{V_2}(\bolds{\rho})
\\
&:= & \biggl[ \frac{H(\bolds{\rho}) - H([\delta,1-\delta]) -
{\delta} \log(M-1)}{I_{\max}(M)}
\\
&&\hspace*{4pt}{} + \frac{\log ((1-L^{-1})/L^{-1}) -
\log((1-\delta)/\delta) - \xi_M}{D_{\max}(M)}
\\
&&\hspace*{100pt}{}\times \mathbf{1}_{\{\max_{i\in\Omega_M} \rho_i \le1 -\delta\}}- K'_2 \biggr]^+,
\end{eqnarray*}
%
%where $K'_2$ is a positive constant independent of $\L$ and $M$.
where $K'_2$ is a constant independent of $\delta$ and ${L}$ whose
closed form is given in the supplemental
article~\cite{Naghshvar-SuppA}, equation~(151).\footnote{As
it will be discussed in Section~\ref{GapLM}, $K'_2$ can be selected
independent of $M$ as well if \mbox{$\sup_M \xi_M < \infty$}.}
%and $I(\bolds{\rho};q^a_{\bolds{\rho}})= H(\bolds{
%is the mutual information between $\theta$ and observation $Z$ under
%action $a$.
%%\label{DefI}
%I(\bolds{\rho};q^a_{\bolds{\rho}})
%: = H(\bolds{\rho}) - \int H({\bolds{\Phi}}^a(\bolds{
\end{theorem}

The proof of Theorem~\ref{thmLB2} is provided in
\hyperref[AppLB]{Appendix}.
%The proof of Proposition~\ref{thmLB2} is based on Lemma~\ref{lemVLB}
%and is provided in Appendix~\ref{AppLB2}.
%The proof of $V^*(\bolds{\rho}) \ge\underline{V_3}(\bolds{

Theorem~\ref{thmLB2} can be used to show that when ${L}< \frac{\log
M}{I_{\max}(M)}$, Problem~\ref{probp} will have a trivial solution. The
precise statement is given by the following corollary.

%
%co1 #&#
\begin{corollary}
\label{PeTo1} Let ${L}< \frac{\log M}{I_{\max}(M)}$, and suppose the
decision maker has a uniform prior belief about the hypotheses. For
sufficiently large $M$, the optimal policy randomly guesses the true
hypothesis without collecting any observation, hence,
$\operatorname{Pe}$, the probability of making a wrong declaration,
approaches $1-\frac{1}{M}$.
\end{corollary}

The proof of Corollary~\ref{PeTo1} is provided in the supplemental
article~\cite{Naghshvar-SuppA}, Section~2.1.
%in Appendix~\ref{AppLB2Cor}.

%
%re2 #&#
\begin{remarks}
The lower bounds in Theorems~\ref{thmLB} and~\ref{thmLB2} can be
explained by the following intuition: for any measure of uncertainty
$U\dvtx \mathbb {P}(\Omega_M)\to\mathbb{R}_+$, the number of samples
required to reduce the uncertainty down to a target level
$U_{\mathrm{target}}$ has to be at least
$\frac{U(\bolds{\rho}(0))-U_{\mathrm{target}}}{\Delta_{\max}(U)}$,
where $\Delta_{\max}(U)$ is the maximum amount of reduction in $U$
associated with a single sample, that is, $\Delta_{\max}(U) =
\max_{a\in\mathcal{A}_M} \max_{\bolds{\rho}\in\mathbb{P}(\Omega_M)}
\{U(\bolds{\rho})-(\mathbb{T}^{a}U)(\bolds{\rho})\}$.
The lower bound in Theorem~\ref{thmLB} is associated with such a lower
bound when taking $U$ to be the log-likelihood function, while the
lower bound in Theorem~\ref{thmLB2} is associated with setting~$U$ to
be the Shannon entropy.
\end{remarks}

%--UPPER
%%%BOUND--------------------------------------------------------------------------
%s4.2 #&#
\subsection{Upper bounds for $V^{*}$}\label{SecUB}

Next we propose two Markov policies $\tilde{\pi}_1$ and $\tilde{\pi}_2$.
Policies $\tilde{\pi}_1$ and $\tilde{\pi}_2$ have two operational phases.
Phase~1 is the phase in which the belief about all hypotheses is below
a certain threshold,
while in phase~2, the belief about one of the hypotheses has passed
that threshold
and actions are selected in favor of that particular hypothesis.
The difference between the two policies is in the actions they take in
each phase.
%The idea of having policies with two phases of operation is not new
%and was applied
%previously in \cite{Burnashev76,Yamamoto79,Caire06} to provide
%achievable bounds for
%the problem of variable-length coding with noiseless feedback.
%This problem is a special case of active hypothesis testing and is
%discussed
%in Subsection~\ref{VLcoding}.
%

%As noted in Section~\ref{ProblemFormulation}, the set of all sensing
%actions is denoted by $\mathcal{A}_M=\{1,2,\ldots,K\}$.
%Let $\Lambda= \{ \bolds{\lambda} \in[0,1]^{K}: \sum_{i=1}^{K}
%
%Let $\bolds{\mu}^*_0 = [\mu^*_{01}, \mu^*_{02}, \ldots,
%$\bolds{\eta}_0 = [\eta_{01}, \eta_{02}, \ldots, \eta_{0K}]$ be
%vectors in $\Lambda$ such that

First we describe policy $\tilde{\pi}_1$. Let $\bolds{\mu}_0$ and
$\bolds{\mu}_i$, $i\in\Omega_M$, be vectors in
$\mathbb{P}(\mathcal{A}_M)$ such that
\begin{eqnarray*}
\bolds{\mu}_0 &:=& \mathop{\argmax}_{\bolds{\lambda} \in
\mathbb{P}(\mathcal{A}_M)} \min_{i\in\Omega_M} \min_{j \neq i} \sum
_{a \in\mathcal{A}_M} \lambda_{a} D\bigl(q^{a}_i
\|q^{a}_j\bigr),
\\
\bolds{\mu}_i &:=& \mathop{\argmax}_{\bolds{\lambda} \in
\mathbb{P}(\mathcal{A}_M)} \min_{j \neq i} \sum
_{a
\in\mathcal{A}_M} \lambda_{a} D\bigl(q^a_i
\|q^a_j\bigr)\qquad \forall i\in\Omega_M.
\end{eqnarray*}
Moreover, let $\mu_{0a}$ and $\mu_{ia}$ denote elements of
$\bolds{\mu}_0$ and $\bolds{\mu}_i$ corresponding to
$a\in\mathcal{A}_M$, respectively. Consider a threshold $\tilde{\rho}$,
$\tilde{\rho}>\frac{1}{2}$.
%$\frac{1}{2}\le\tilde{\rho}\le\max\{\frac{1}{2},1-L^{-1}\}$. %$\tilde{
Markov (randomized) policy $\tilde{\pi}_1$ is defined as
follows:\footnote{Policies $\tilde{\pi}_1$ and $\tilde{\pi}_2$ are
not unique; they each represent a class of parameterized policies. In
fact, the tilde in $\tilde{\pi}_1$ and $\tilde{\pi}_2$ has been
chosen to emphasize the dependency of these policies on the
threshold/parameter $\tilde{\rho}$.}
\begin{itemize}
\item If $\rho_i \ge1-{L}^{-1}$, retire and select $H_i$ as the true
hypothesis.

\item If $\rho_i \in[\tilde{\rho},1-{L}^{-1})$, then
\begin{itemize}
\item $\tilde{\pi}_1(a|\bolds{\rho})= \mu_{ia}$ $\forall a
    \in\mathcal{A}_M$.
% \item$\tilde{\pi}_2(a|\bolds{\rho}) = \eta_{ia}$ $\forall a \in
\end{itemize}
%
% \item If $\rho_i \in[0,\tilde{\rho})$ for all $i\in\Omega_M$, then
%
\item If $\rho_i < \min\{\tilde{\rho},1-L^{-1}\}$ for all $i\in
\Omega_M$, then
\begin{itemize}
\item $\tilde{\pi}_1(a|\bolds{\rho}) = \mu_{0a}$ $\forall a
    \in\mathcal{A}_M$.
% \item$\tilde{\pi}_2(a|\bolds{\rho}) = \eta_{0a}$ $\forall a \in
\end{itemize}
%
% where $\mu_{0a}$ and $\mu_{ia}$ represent elements of $\bolds{
% corresponding to $a\in\mathcal{A}_M$, respectively.
\end{itemize}

In \cite{Chernoff59}, Chernoff proposed a policy that, at each time
$t$, selects action~$a$ with probability $\mu_{i^*a}$, where
$i^*=\argmax_{i\in\Omega_M}\rho_i(t)$ denotes the most likely true
hypothesis. In other words, $\tilde{\pi}_1$ coincides with Chernoff's
scheme in its second phase and ensures its asymptotic optimality in
${L}$, while its first phase in a very natural manner relaxes the
technical assumption in~\cite{Chernoff59} where all actions were
required to discriminate between all hypotheses pairs. Following
Chernoff's approach (Theorem~1 in \cite{Chernoff59}), we can analyze the
performance of policy $\tilde{\pi}_1$ and obtain the following upper
bound for $V^*$.

For notational simplicity, let
\begin{eqnarray*}
I_{\bolds{\mu}_0}(M) &:=& \min_{i\in\Omega_M} \min_{j \neq i} \sum_{a \in\mathcal{A}_M}
\mu_{0a} D\bigl(q^{a}_i\|q^{a}_j
\bigr),
\\
%I_{\bolds{\mu},\tilde{\rho}}(M) &:= \min_{i\in\Omega_M}
%D(q^{a}_i\|q^{a}_j),\\
%I_1(M) &:= \left(\frac{\log M + \xi_M}{I_{\bolds{\mu},\tilde{
I_1(M)
&:=& \biggl(\frac{\log M + 4\xi_M}{\min_{i\in\Omega
_M} \min_{j \neq i} \sum_{a \in\mathcal{A}_M} \mu
_{ja} D(q^{a}_i\|q^{a}_j)} \biggr)^{-2} I_{\bolds{\mu}_0}(M),
\\
%I_1(M) &:= \max_{\bolds{\lambda} \in\mathbb{P}(
%I_2 &:= \max_{\bolds{\lambda} \in\mathbb{P}(
%
D_{\bolds{\mu}_i}(M) &:=& \min_{j \neq i} \sum
_{a \in\mathcal{A}_M} \mu_{ia} D\bigl(q^a_i
\|q^a_j\bigr)\qquad \forall i\in \Omega_M.
%D_{\bolds{\eta}_i} &:= \max_{\bolds{\lambda} \in
\end{eqnarray*}

%
%pr2 #&#
\begin{proposition}
\label{propUB1} Under Assumptions~\ref{KL0} and~\ref{Jump}, and for
${L}> 1$, $\bolds{\rho} \in\mathbb{P}_{{L}}(\Omega_M)$, and arbitrary
$\iota\in(0,1)$,
\begin{eqnarray*}
V^*(\bolds{\rho}) &\le&\widebar{V}_1(\bolds{\rho})
\\
&:=&
\frac{H(\bolds{\rho}) + \log M + \log(\tilde{\rho
}/(1-\tilde{\rho}))}{I_1(M)}(1+\iota) + \sum_{i=1}^{M}
\rho_i\frac{\log{L}}{D_{\bolds{\mu
}_i}(M)}(1+\iota)
\\
&&{} + M \biggl(2+\frac{1}{((\iota/2)/(1+\iota))^5 (I_1(M)/(2\xi_M))^4} \biggr)
\\
&&\hspace*{10pt}{}\times \Bigl(L\Bigl(1-\max
_{j\in\Omega_M} \rho_j\Bigr) \Bigr)^{-(\iota^3/(1+\iota)^2) I^2_1(M)/(4\xi^3_M)} +2.
\end{eqnarray*}
%
%is an upper bound for the optimal value function $V^{*}$.
\end{proposition}

The proof is based on a performance analysis of policy $\tilde{\pi }_1$
and is provided in the supplemental article~\cite{Naghshvar-SuppA},
Section~5.2. %\ref{supp:UB1}.

%Under Assumptions~\ref{KL0} and~\ref{Jump}, and for $\L> 1$ and any $
%%\frac{H(\bolds{\rho}) + \log(M-1)}{I_1} +
%_{k \neq i} \log\frac{\rho_i}{\rho_k}}{D_{\bolds{
%is an upper bound for the optimal value function $V^{*}$
%where $\frac{o(M + \log\L)}{M + \log\L} \to0$ as $M \L\to\infty$.
%
%%\begin{pf}
%The proof is done by analyzing the performance of policy $\tilde{
%%\end{pf}

Next we describe policy $\tilde{\pi}_2$. Let $\bolds{\eta}_0$ and
$\bolds{\eta}_i$, $i\in\Omega_M$, be vectors in
$\mathbb{P}(\mathcal{A}_M)$ such that
\begin{eqnarray*}
\bolds{\eta}_0 &:=& \mathop{\argmax}_{\bolds{\lambda} \in
\mathbb{P}(\mathcal{A}_M)} \min_{i\in\Omega_M} \min_{{\hat{\bolds{\rho}}}\in\mathbb{P}_{{L}}(\Omega_M)} \sum
_{a \in\mathcal{A}_M} \lambda_{a} D\biggl(q^{a}_i
\bigg\| \sum_{j\neq i} \frac{{\hat{\rho}}_j}{1-\hat{\rho}_i} q^{a}_j
\biggr),
\\
\bolds{\eta}_i &:=& \mathop{\argmax}_{\bolds{\lambda} \in
\mathbb{P}(\mathcal{A}_M)} \min_{{\hat{\bolds{\rho}}}\in\mathbb{P}_{{L}}(\Omega_M)} \sum
_{a \in\mathcal
{A}_M} \lambda_{a} D\biggl(q^{a}_i
\bigg\| \sum_{j\neq i} \frac{{\hat{\rho
}}_j}{1-\hat{\rho}_i} q^{a}_j
\biggr)\qquad \forall i\in\Omega_M.
\end{eqnarray*}
Moreover, let $\eta_{0a}$ and $\eta_{ia}$ denote elements of
$\bolds{\eta}_0$ and $\bolds{\eta}_i$ corresponding to
$a\in\mathcal{A}_M$, respectively. Consider a threshold $\tilde{\rho}$,
$\tilde{\rho}>\frac{1}{2}$.
%$\frac{1}{2}\le\tilde{\rho}\le\max\{\frac{1}{2},1-L^{-1}\}$. %$\tilde{
Markov (randomized) policy $\tilde{\pi}_2$ is defined as follows:
\begin{itemize}
\item If $\rho_i \ge1-{L}^{-1}$, retire and select $H_i$ as the true
hypothesis.
\item If $\rho_i \in[\tilde{\rho},1-{L}^{-1})$, then
\begin{itemize}
% \item$\tilde{\pi}_1(a|\bolds{\rho})= \mu_{ia}$ $\forall a \in
%
\item$\tilde{\pi}_2(a|\bolds{\rho}) = \eta_{ia}\ \forall a
    \in\mathcal{A}_M$.
\end{itemize}
%
% \item If $\rho_i \in[0,\tilde{\rho})$ for all $i\in\Omega_M$, then
%
\item If $\rho_i < \min\{\tilde{\rho},1-L^{-1}\}$ for all $i\in
\Omega_M$, then
\begin{itemize}
% \item$\tilde{\pi}_1(a|\bolds{\rho}) = \mu_{0a}$ $\forall a \in
%
\item$\tilde{\pi}_2(a|\bolds{\rho}) = \eta_{0a}\ \forall a
    \in\mathcal{A}_M$.
\end{itemize}
\end{itemize}

For notational simplicity, let
\begin{eqnarray*}
%I_1 &:= \max_{\bolds{\lambda} \in\mathbb{P}(
%I_2(M) &:= \max_{\bolds{\lambda} \in\mathbb{P}(
I_{\bolds{\eta}_0}(M) &:=& \min_{i\in\Omega_M}\ \min_{{\hat{\bolds{\rho}}}\in\mathbb{P}_{{L}}(\Omega_M)} \sum
_{a \in\mathcal{A}_M} \eta_{0a} D\biggl(q^{a}_i
\bigg\| \sum_{j\neq i} \frac{{\hat{\rho}}_j}{1-\hat{\rho}_i} q^{a}_j
\biggr),
\\
I_{\bolds{\eta},\tilde{\rho}}(M) &:=& \min_{i\in
\Omega_M} \min_{k\neq i} \min_{{\hat{\bolds{\rho}}}\dvtx \hat{\rho}_k\ge\tilde{\rho}} \sum_{a \in
\mathcal{A}_M}
\eta_{ka} D\biggl(q^{a}_i\bigg\| \sum
_{j\neq i} \frac{{\hat
{\rho}}_j}{1-\hat{\rho}_i} q^{a}_j
\biggr),
\\
I_2(M) &:=& \min \bigl\{ I_{\bolds{\eta}_0}(M), I_{\bolds{\eta},\tilde{\rho}}(M)
\bigr\},
\\
%
%D_{\bolds{\mu}_i} &:= \max_{\bolds{\lambda} \in
D_{\bolds{\eta}_i}(M) &:=& \min_{{\hat{\bolds{\rho}}}\in\mathbb{P}_{{L}}(\Omega_M)}
\sum_{a \in
\mathcal{A}_M} \eta_{ia} D\biggl(q^{a}_i
\bigg\| \sum_{j\neq i} \frac{{\hat
{\rho}}_j}{1-\hat{\rho}_i} q^{a}_j
\biggr)\qquad \forall i\in\Omega_M.
\end{eqnarray*}
%
%It can be easily shown that $I_2$ and $D_{\bolds{\eta}_i}$, $
%
%Note that Table~\ref{ListNote1} provides a list of the notations
%introduced in this section.
%Note that Tables~\ref{ListNote1} and~\ref{ListNote2} provide
%respectively
%a list of the notations introduced in this section and their limiting
%values.
%Note that Table~\ref{ListNote1} provides a list of the notations
%introduced in this section.

%Propositions~\ref{thmUB2} at the end of this section provides
%an upper bound $\widebar{V}$ for the value function $V^*$.

%
%Under Assumptions~\ref{KL0} and~\ref{Jump}, and for $\L> 1$ and any $
%%\frac{H(\bolds{\rho}) + \log(M-1)}{I_1} +
%_{k \neq i} \log\frac{\rho_i}{\rho_k}}{D_{\bolds{\mu}_i}}
%+ \frac{o(M + \log\L)}{I_1^2},
%is an upper bound for the optimal value function $V^{*}$
%where $\frac{o(M + \log\L)}{M + \log\L} \to0$ as $M \L\to\infty$.
%
%%\begin{pf}
%The proof is done by analyzing the performance of policy $\tilde{
%%\end{pf}

%
%th3 #&#
\begin{theorem}
\label{thmUB2} Under Assumptions~\ref{KL0} and~\ref{Jump}, and for
${L}> 1$ and any $\bolds{\rho} \in\mathbb{P}_{{L}}(\Omega_M)$,
\[
V^*(\bolds{\rho}) \le\widebar{V}_2 (\bolds{\rho}):= \frac{H(\bolds{\rho})+\log (\tilde{\rho}/(1-\tilde
{\rho})) + \xi_M + \log e}{I_2(M)} + %\sum_{i=1}^M \rho_i \frac{\log\frac{1-\L^{-1}}{\L^{-1}}}{D_{
\sum_{i=1}^M \rho_i
\frac{\log{L}}{D_{\bolds{\eta}_i}(M)} + 1.
\]
%
%is an upper bound for the optimal value function $V^{*}$.
\end{theorem}

The proof is based on a performance analysis of policy $\tilde{\pi }_2$
and is provided in the \hyperref[AppLB]{Appendix}.

%Policies $\tilde{\pi}_1$ and $\tilde{\pi}_2$ are equivalent for $M=2$.
%
%The performance of policies $\tilde{\pi}_1$ and $\tilde{\pi}_2$ depend
%highly on the problem at hand.
%Sections~\ref{Gap} and~\ref{Rate} will elaborate on this.
%%In general, policy $\tilde{\pi}_2$ is more efficient when: 1) $M$ is
%large; and
%%2) for each hypothesis $H_i$, $i\in\Omega_M$, there exists an action
%$a\in\mathcal{A}_M$
%%that can efficiently distinguish $H_i$ from the combination of other
%hypotheses.
%%While $\tilde{\pi}_1$ performs better when 1) $M$ is small; and
%%2) there is no action that can efficiently distinguish one hypothesis
%from the combination of other hypotheses.

%{\scriptsize{
% \hline
% Notation & \hspace{.5in} Description \\
% \hline
% \vspace{0.05 in}
% $I_{\max}(M)$ & $\max_{a \in\mathcal{A}_M} \max
%_{\hat{\bolds{\rho}}\in\mathbb{P}(\Omega_M)}
%I(\hat{\bolds{\rho}};q^{a}_{\hat{\bolds{\rho}}})$ \\
% \hline
% \vspace{0.05 in}
% $D_{\max}(M)$ & $\max_{i,j\in\Omega_M} \max
%_{a \in\mathcal{A}_M} D(q_i^{a}\|q_j^{a})$ \\
% \hline
% \vspace{0.05 in}
% $I_1(M)$ & $\max_{\bolds{\lambda} \in\mathbb{P}(
% \hline
% \vspace{0.05 in}
% $I_2(M)$ & $\max_{\bolds{\lambda} \in\mathbb{P}(
% \hline
% \vspace{0.05 in}
% $D_{\bolds{\mu}_i}(M)$ & $\max_{\bolds{\lambda} \in
% \hline
% \vspace{0.05 in}
% $D_{\bolds{\eta}_i}(M)$ & $\max_{\bolds{\lambda}
% \hline
%}}

%
%t1 #&#
\begin{table}
\tabcolsep=0pt
 \caption{Summary of notation} \label{ListNote1}
\begin{tabular*}{\textwidth}{@{\extracolsep{\fill}}@{}ll@{}}
\hline
\textbf{Notation} & \multicolumn{1}{c@{}}{\textbf{Description}} \\
\hline
$I_{\max}(M)$ & $\max_{a \in\mathcal{A}_M} \max
_{\hat{\bolds{\rho}}\in\mathbb{P}(\Omega_M)}
I(\hat{\bolds{\rho}};q^{a}_{\hat{\bolds{\rho}}})$
\\[8pt]
$D_{\max}(M)$ & $\max_{i,j\in\Omega_M} \max_{a \in\mathcal{A}_M} D(q_i^{a}\|q_j^{a})$
\\[8pt]
$\bolds{\mu}_0$ & $\argmax_{\bolds{\lambda} \in
\mathbb{P}(\mathcal{A}_M)} \min_{i\in\Omega_M} \min_{j \neq i} \sum_{a \in\mathcal{A}_M} \lambda_{a}
D(q^{a}_i\|q^{a}_j)$
\\[8pt]
$\bolds{\mu}_i$ & $\argmax_{\bolds{\lambda} \in
\mathbb{P}(\mathcal{A}_M)} \min_{j \neq i} \sum_{a
\in\mathcal{A}_M} \lambda_{a} D(q^a_i\|q^a_j)$
\\[8pt]
$I_{\bolds{\mu}_0}(M)$ & $\min_{i\in\Omega_M} \min_{j \neq i} \sum_{a \in\mathcal{A}_M} \mu_{0a}
D(q^{a}_i\|q^{a}_j)$
\\[8pt]
%
% $I_{\bolds{\mu},\tilde{\rho}}(M)$ & $\min_{i\in\Omega_M}
%D(q^{a}_i\|q^{a}_j)$ \\[8pt]
% $I_1(M)$ & $\left(\frac{\log M + \xi_M}{I_{\bolds{\mu},\tilde{
$I_1(M)$ & $ (\frac{\log M + 4\xi_M}{\min_{i\in\Omega
_M} \min_{j \neq i} \sum_{a \in\mathcal{A}_M} \mu
_{ja} D(q^{a}_i\|q^{a}_j)} )^{-2} I_{\bolds{\mu}_0}(M)$
\\[8pt]
$D_{\bolds{\mu}_i}(M)$ & $\min_{j \neq i} \sum_{a \in\mathcal{A}_M} \mu_{ia} D(q^a_i\|q^a_j)$
\\[8pt]
$\bolds{\eta}_0$ & $\argmax_{\bolds{\lambda} \in
\mathbb{P}(\mathcal{A}_M)} \min_{i\in\Omega_M} \min_{{\hat{\bolds{\rho}}}\in\mathbb{P}_{{L}}(\Omega_M)}
\sum_{a \in\mathcal{A}_M} \lambda_{a} D(q^{a}_i\| \sum_{j\neq i} \frac{{\hat{\rho}}_j}{1-\hat{\rho}_i} q^{a}_j)$
\\[8pt]
$\bolds{\eta}_i$ & $\argmax_{\bolds{\lambda} \in
\mathbb{P}(\mathcal{A}_M)} \min_{{\hat{\bolds{\rho}}}\in\mathbb{P}_{{L}}(\Omega_M)} \sum_{a \in\mathcal
{A}_M} \lambda_{a} D(q^{a}_i\| \sum_{j\neq i} \frac{{\hat{\rho
}}_j}{1-\hat{\rho}_i} q^{a}_j)$
\\[8pt]
$I_{\bolds{\eta}_0}(M)$ & $\min_{i\in\Omega_M} \min_{{\hat{\bolds{\rho}}}\in\mathbb{P}_{{L}}(\Omega_M)}
\sum_{a \in\mathcal{A}_M} \eta_{0a} D(q^{a}_i\| \sum_{j\neq i} \frac{{\hat{\rho}}_j}{1-\hat{\rho}_i} q^{a}_j)$
\\[8pt]
$I_{\bolds{\eta},\tilde{\rho}}(M)$ & $\min_{i\in
\Omega_M} \min_{k\neq i} \min_{{\hat{\bolds{\rho}}}\dvtx \hat{\rho}_k\ge\tilde{\rho}} \sum_{a \in
\mathcal{A}_M} \eta_{ka} D(q^{a}_i\| \sum_{j\neq i} \frac{{\hat
{\rho}}_j}{1-\hat{\rho}_i} q^{a}_j)$
\\[8pt]
$I_2(M)$ & $\min \{ I_{\bolds{\eta}_0}(M), I_{\bolds{\eta},\tilde{\rho}}(M)  \}$
\\[8pt]
$D_{\bolds{\eta}_i}(M)$ & $\min_{{\hat{\bolds{\rho}}}\in\mathbb{P}_{{L}}(\Omega_M)} \sum_{a \in
\mathcal{A}_M} \eta_{ia} D(q^{a}_i\| \sum_{j \neq i} \frac{{\hat
{\rho}}_j}{1-\hat{\rho}_i} q^{a}_j)$ \\
\hline
\end{tabular*}
\end{table}

%%%%%%%%%%%%%%%%%%%%%%%%%%%%%%%%%%%%%%%%%%%%%%%%%%%%%%%%
%%% Applications and Techinal Consequences of the Bounds
%%%%%%%%%%%%%%%%%%%%%%%%%%%%%%%%%%%%%%%%%%%%%%%%%%%%%%%%
%s5 #&#
\section{Asymptotic analysis and consequences}\label{Optimality}

In this section we state and discuss the consequence of the bounds
obtained in Section~\ref{SecB} in asymptotically large ${L}$ and $M$.
Note that Table~\ref{ListNote1} provides a list of the notation
introduced in Section~\ref{SecB}.
%It can be easily shown that $I_2(M) \le D_{\bolds{\eta}_i}(M) \le
%D_{\bolds{\mu}_i}(M) \le D_{\max}(M)$, $\forall i\in
%Furthermore, let $\underline{I}_2:=\inf_M I_2(M)$, and $\underline{D}_{
%Here we note that the analysis above is a function of the hypothesis
%set $\Omega_M$,
%hence number of hypotheses, $M$.
%Because of our interest to study the asymptotically large hypothesis
%set,
%we stop suppressing variable $M$, i.e., let $\Omega_M:=\{1,2,\ldots,M
%Furthermore, we emphasize this dependence in Table~\ref{ListNote1}.
%Note that, by definition in Table~\ref{ListNote1}, $I_{
%%First, we focus on the advantage of adaptively selecting sensing
%actions.
%%In particular, we show that the performance gap between the best
%(randomized) non-adaptive policy
%%and $\tilde{\pi}_1$ (hence the optimal one) grows
%%at least logarithmically as the penalty $\L$ increases.
%%In Subsection~\ref{Gap}, we study the gap between lower and upper
%bounds provided in Subsection~\ref{SecB}
%%and discuss the asymptotic optimality of the proposed heuristics $

%%%%%%%%%%%%%%%%%%%%%%%%%%%%%%%%%%%%%%%%%%%%%
%Asymptotic Optimality
%%%%%%%%%%%%%%%%%%%%%%%%%%%%%%%%%%%%%%%%%%%%%
%s5.1 #&#
\subsection{Order and asymptotic optimality in ${L}$}\label{GapL}

The lower and upper bounds provided in Section~\ref{SecB} %by Propositions~\ref{LemmaLB} and~\ref{LemmaUB}
can be applied to establish the order optimality and asymptotic
optimality %(see Definition~\ref{AsymDef})
of the proposed policies as defined below. Let $V_{\pi}(\bolds{\rho})$
denote the value function for policy $\pi$, that is, the expected total
cost achieved by policy $\pi$ when the initial belief
is~$\bolds{\rho}$.
%
%problematic as $M\to\infty$.
%As $M\to\infty$, the state space $\mathbb{P}(\Omega_M)$ also changes
%and keeping the prior $\bolds{\rho}$ does not make sense.
%We probably should define asymptotic optimality in L and M only for
%uniform priors.}

\begin{definition}
\label{OrdDef} For fixed $M$, policy $\pi$ is referred to as
\textit{order optimal} in ${L}$ if for all $\bolds{\rho}
\in\mathbb{P}(\Omega_M)$,
\[
\lim_{{L}\to\infty} \frac{V_{\pi}(\bolds{\rho}) -
V^*(\bolds{\rho})}{V_{\pi}(\bolds{\rho})}<1.
\]
\end{definition}

\begin{definition}
\label{AsymDef}
%Let $V_{\pi}(\bolds{\rho})$ denote the value function for policy $
%the expected total cost achieved by policy $\pi$ when the initial
%belief is~$\bolds{\rho}$.
For fixed $M$, policy $\pi$ is referred to as \textit{asymptotically
optimal of order}-1 in ${L}$ if for all $\bolds{\rho} \in
\mathbb{P}(\Omega_M)$,
\[
\lim_{{L}\to\infty} \frac{V_{\pi}(\bolds{\rho}) -
V^*(\bolds{\rho})}{V_{\pi}(\bolds{\rho})}=0.
\]
\end{definition}

\begin{definition}
\label{Asym2Def} For fixed $M$, policy $\pi$ is referred to as
\textit{asymptotically optimal of order}-2 in ${L}$ if for all
$\bolds{\rho} \in \mathbb{P}(\Omega_M)$, there exists a constant $B$
independent of ${L}$ such that
\[
V_{\pi}(\bolds{\rho}) - V^*(\bolds{\rho}) \le B.
\]
\end{definition}
%
%Note that the bounds provided by Propositions~\ref{LemmaLB} and~
%are logarithmic in $\L$ and $M$. % (this already establishes that the
%proposed policies are \emph{order optimal}).
%In order to establish asymptotic optimality of the proposed policies,
%it suffices to show that
%the gap between the lower and upper bounds grows sub-logarithmically
%as $\L$ or $M$ increases %(or equivalently with increasing number of
%samples),
%and hence, the gap grows in a much slower rate than the bounds
%themselves.
%The next two lemmas provide sufficient conditions under which $\tilde{
%are asymptotic optimal.
%

%
%re3 #&#
\begin{remarks}
It is clear from the definitions above that order optimality is weaker
than asymptotic optimality of order-1, while asymptotic optimality of
order-2 is the strongest notion. The notion of asymptotic optimality of
order-1 was first introduced in~\cite{Chernoff59},
which naturally motivates the extension of higher orders. % proposed in
%It is clear from the definitions above that order optimality is weaker
%than asymptotic optimality of order-1.
%If policy $\pi$ is asymptotically optimal of order-1 in $\L$ (and $M$),
%then $V_{\pi}$ and $V^*$ will have the same dominating terms in $\L$
%(and $M$);
%while order optimality of policy $\pi$ only implies that
%dominating terms in $V_{\pi}$ and $V^*$ are the same up to a constant
%factor.
%The notion of asymptotic optimality of order-1 was first introduced in
%%
%There exist stronger notions of asymptotic optimality in the
%literature \cite{Moustakides11}
%such as asymptotic optimality of order-2 as defined above which
%implies that
%the difference between $V_{\pi}$ and $V^*$ is bounded.
%
%or the difference goes to zero as $\L$ goes to infinity (asymptotic
%optimality of order-3).
%Although our lower and upper bounds allow us to establish asymptotic
%optimality as of order-2
%for some special cases of active hypothesis testing (see Theorem~
%they do not provide such results in general.
\end{remarks}

%{\scriptsize{
%{\caption{Summary of limiting values}
% \hline
% Notation & Description & Notation & Description \\
%% \hline
%% \vspace{0.05 in}
%% $\xi(M)$ & $\max_{i,j\in\Omega_M(M)} \max_{a \in
% \hline
% \vspace{0.05 in}
% $\underline{I}_{\max}$ & $I_{\max}(2)$ & $
% \hline
% \vspace{0.05 in}
% $\underline{D}_{\max}$ & $D_{\max}(2)$ & $
%% \hline
%% \vspace{0.05 in}
%% $\underline{I}_1$ & $\inf_{M} I_1$ & $\widebar{I}_1$ & $
% \hline
% \vspace{0.05 in}
% $\underline{I}_2$ & $\inf_{M} I_2(M)=\lim_{M\to
% \hline
%}}

The next corollary establishes order and asymptotic optimality of our
proposed policies.

%Policy $\tilde{\pi}_1$ is asymptotically optimal of order-1 in $\L$.
%%
%The proof simply follows from Definition~\ref{AsymDef}, Fact~
%%\begin{eqnarray}
%%\label{pi1asym02}
%%\min_{j \neq i} \max_{{a}\in\mathcal{A}_M} D(q^{a}_i\|q^{a}_j)
%%\ge\max_{\bolds{\lambda} \in\mathbb{P}(\mathcal{A}_M) }
%%\ge\max_{{a} \in\mathcal{A}_M} \min_{j \neq i} D(q^{a}_i \|
%q^{a}_j).
%%\end{eqnarray}

%%\label{ThmAsym1}
%Policy $\tilde{\pi}$ is asymptotically optimal of order-2 in $\L$ if
%
%Note that Theorem~\ref{} can be compared with \cite{}
%where the proposed policy can be shown to be always asymptotic optimal
%of order-1 in $\L$.
%In contrast, $\tilde{\pi}$ is shown to be asymptotic optimal of
%order-2 in $\L$ only in cases where \eqref{} holds.
%}

%
%co2 #&#
\begin{corollary}
\label{corAsym1} Under Assumptions~\ref{KL0} and \ref{Jump}, policy
$\tilde{\pi}_1$ is asymptotically optimal of order-1 in ${L}$.
Furthermore, policy $\tilde{\pi}_2$ attains asymptotic optimality of
\mbox{order-2} in ${L}$ if
%e7 #&#
\begin{equation}
\label{pi2asym01} \min_{j \neq i} \max_{a \in\mathcal{A}_M} D
\bigl(q^{a}_i\|q^{a}_j\bigr) =
D_{\bolds{\eta}_i}(M)\qquad \forall i\in\Omega_M.
\end{equation}
\end{corollary}
\begin{pf}
%The proof simply follows from Definition~\ref{AsymDef}, Fact~
Using Proposition~\ref{propLB1} and by setting $\delta=(\log
{L})^{-1/3}$, we obtain
%e8 #&#
\begin{equation}
\label{LB1asym} V^*(\bolds{\rho})\ge\sum_{i=1}^M
\rho_i \frac{\log
{L}}{D_{\mu_i}(M)}+O \bigl((\log{L})^{2/3} \bigr).
\end{equation}
On the other hand, from Proposition~\ref{propUB1} and by setting
$\iota=(\log{L})^{-1/4}$, we get
%e9 #&#
\begin{equation}
\label{UB1asym} V^*(\bolds{\rho})\le\sum_{i=1}^M
\rho_i \frac{\log
{L}}{D_{\mu_i}(M)}+O \bigl((\log{L})^{3/4} \bigr).
\end{equation}
The proof of the first part of the corollary simply follows from
Definition~\ref{AsymDef}, inequality~(\ref{LB1asym})
and~(\ref{UB1asym}).

Similarly, the proof of the second part of the corollary follows from
Definition~\ref{Asym2Def}, Theorems \ref{thmLB}~and~\ref{thmUB2}.
\end{pf}

\subsection{Order and asymptotic optimality in both ${L}$ and $M$}\label{GapLM}
As mentioned in Section~\ref{sum}, one of the main drawbacks of
Chernoff's asymptotic optimality notion was his neglecting the
complementary role of parameter $M$. In particular, the notion of
asymptotic optimality in ${L}$ falls short in showing the tension
between using an (asymptotically) large number of samples to
discriminate among a few hypotheses with (asymptotically) high accuracy
or an (asymptotically)\vadjust{\goodbreak} large number of hypotheses with a lower degree
of accuracy. In this section we address this issue by analyzing the
bounds when ${L}$ and $M$ are both asymptotically large. More
specifically, we consider a sequence of problems indexed by parameter
$M$ in which the set of actions and observation kernels grow
monotonically as $M$ increases, that is, for all $M<M'$,
%e10 #&#
\begin{equation}\label{MonotonicGrowth1}
\mathcal{A}_M\subseteq\mathcal{A}_{M'}\quad\mbox{and}\quad
\bigl\{ q^a_i(\cdot)\bigr\}_{i\in\Omega_M,a\in\mathcal{A}_M}
\subseteq\bigl\{ q^a_i(\cdot)\bigr\}_{i\in\Omega_{M'},a\in\mathcal{A}_{M'}}.
\end{equation}

Recall the notation listed in Table~\ref{ListNote1}. Also, let $D_1(M)$
and $D_2(M)$ denote, respectively, the harmonic mean of $\{
D_{\bolds{\mu}_i}(M) \}_{i\in\Omega_M}$ and $\{ D_{\bolds{\eta}_i}(M)
\}_{i\in\Omega_M}$, that is,
%
%e11 #&#
\begin{equation}
D_1(M) = M \Biggl(\sum_{i=1}^M
\frac{1}{D_{\bolds{\mu}_i}(M)} \Biggr)^{-1},\qquad D_2(M) = M \Biggl(
\sum_{i=1}^M \frac{1}{D_{\bolds{\eta}_i}(M)}
\Biggr)^{-1}.
\end{equation}
Moreover, let
%
%e12 #&#
%e13 #&#
%e14 #&#
\begin{eqnarray}
\widebar{I}_{\max} &:=& \sup_{M} I_{\max}(M),\qquad
\widebar{D}_{\max}:= \sup_{M}
D_{\max}(M),
\\
\underline{I}_{\max} &:=& \inf_{M}
I_{\max}(M),\qquad \underline{D}_{\max}:= \inf
_{M} D_{\max}(M),
\\
\underline{I}_2 &:=&\inf_M I_2(M),\qquad \underline{D}_2:= \inf_{M}
D_2(M).
\end{eqnarray}

By the definition and from~(\ref{MonotonicGrowth1}), $D_{\max}(M)$ and
$I_{\max}(M)$ are nondecreasing in~$M$. Furthermore, from Jensen's
inequality,
%Furthermore, by Jensen's inequality and Assumption~\ref{Jump}, $I_{
%in Section~\ref{WeakenAsmp}.}}
%e15 #&#
\begin{eqnarray}
\label{IlessD}
I_{\max}(M)&=&\max_{a \in\mathcal{A}_M}\, \max_{\hat{\bolds{\rho}}\in\mathbb{P}(\Omega_M)} \sum
_{i=1}^M \hat{\rho}_i D \Biggl(q^a_i\Bigg\|\sum_{j=1}^M \hat{\rho}_j q^a_j\Biggr)\nonumber
\\
&\le&\max_{a \in\mathcal{A}_M}\, \max_{\hat
{\bolds{\rho}}\in\mathbb{P}(\Omega_M)} \sum_{i=1}^M
\hat {\rho}_i \sum_{j=1}^M
\hat{\rho}_j D\bigl(q^a_i
\|q^a_j\bigr)
\\
&\le&\max_{a \in\mathcal{A}_M}\, \max_{i,j\in\Omega
_M} D\bigl(q_i^{a}
\|q_j^{a}\bigr) = D_{\max}(M)\nonumber
\end{eqnarray}
and by Assumption~\ref{Jump}, we have\footnote{Inequality
(\ref{DlessXi}) holds true even if Assumption~\ref{Jump} is replaced by
a more general assumption such as those suggested in
Section~\ref{WeakenAsmp}.}
%e16 #&#
\begin{equation}
\label{DlessXi} D_{\max}(M) \le\max_{i,j\in\Omega_M} \max_{a \in\mathcal{A}_M} \sup
_{z\in\mathcal{Z}} \log\frac
{q^a_i(z)}{q^a_j(z)} \le\xi_M.
\end{equation}
Similarly, $I_2(M) \le D_{\bolds{\eta}_i}(M) \le D_{\bolds{\mu}_i}(M)
\le D_{\max}(M) \le\xi_M$, $\forall i\in\Omega _M$, for all $M$.
%We denote by $\underline{D}_{\max}$ and $\underline{I}_{
%at $M=2$ respectively;
%and by $\widebar{D}_{\max}$ and $\widebar{I}_{\max}$
%their limits as $M\to\infty$.
Since $D_{\max}(M)$ and $I_{\max}(M)$ are nondecreasing in $M$, we have
$\underline{D}_{\max} = D_{\max}(2)$, $\widebar{D}_{\max} =
\lim_{M\to\infty} D_{\max}(M)$, $\underline{I}_{\max} = I_{\max}(2)$ and $\widebar{I}_{\max} =\break \lim_{M\to \infty} I_{\max}(M)$.

%Furthermore, let $\underline{I}_2:=\inf_M I_2(M)$, and $\underline{D}_{

Furthermore, to ensure that the distance between the observation
kernels remains bounded as $M$ increases (and
$\widebar{D}_{\max}<\infty$), we consider the following assumption:

%
%as3 #&#
\begin{assumption}\label{JumpM}
There exists $\xi< \infty$ such that
\[
\sup_M \xi_M \le\xi.
\]
%
% < \infty$.
\end{assumption}

%%For $\L> 1$, $\L> \frac{\log M}{I_{\max}}$, and $\delta=
%> \max\{1, \frac{\log M}{I_{\max}(M)}\}$ and $\delta=\frac{1}{
%{\mathbf{1}}_{\{\max_{i\in\Omega_M} \rho_i \le1 -

This assumption allows us to specialize Theorem~\ref{thmLB2} as follows.

%
%co3 #&#
\begin{corollary}
\label{CorLB2} Let $\bolds{\rho}_{u,M}$ denote a uniform prior on the
set of hypotheses $\Omega_M$.
%For $\L> 1$, $\L> \frac{\log M}{I_{\max}}$, and $\delta=
Under Assumptions~\ref{KL0},~\ref{Jump} and~\ref{JumpM}, and for
$\delta=\frac{1}{\log2 M {L}}$ and ${L}> \max\{2, \frac{\log
M}{I_{\max}(M)}\}$,
\begin{eqnarray*}
\underline{V_2}(\bolds{\rho}_{u,M})&\ge& \biggl[
\frac{\log M - 2}{\widebar{I}_{\max}} + \frac{\log
((1-{L}^{-1})/L^{-1})}{\widebar{D}_{\max}} %-\frac{2}{\underline{I}_{\max}
- \frac{\log\log{L}M+\xi}{\underline{D}_{\max}} -
K'_2 \biggr]^+, %\left[ \frac{H(\bolds{\rho})}{\widebar{I}_{\max}} +
%{\mathbf{1}}_{\{\max_{i\in\Omega_M} \rho_i \le1 -
\end{eqnarray*}
where $K'_2$ is a positive constant independent of ${L}$ and $M$.
%whose closed-form is given in \eqref{K2closeM} in the supplemental
%article~\cite{Naghshvar-SuppA}, Section~\ref{supp:Claims},.
\end{corollary}

The proof of Corollary~\ref{CorLB2} is provided in the supplemental
article~\cite{Naghshvar-SuppA}, Section~2.2. %\ref{supp:CorLB2}.

The next definition extends the notions of order and asymptotic
optimality defined in Section~\ref{GapL} to the case where $M$
increases as well.
\begin{definition}
\label{AsymM}
%Let $\bolds{\rho}_{u,M}$ denote a uniform prior on the set of
%hypotheses $\Omega_M$.
Policy $\pi$ is referred to as \textit{order optimal} and
\textit{asymptotically optimal of order}-1 in ${L}$ and $M$ if,
respectively,\footnote{Note that unlike Definitions~\ref{OrdDef}--\ref{Asym2Def} where we considered the performance gap
between policy $\pi$ and the optimal policy $\pi^*$ for all values of
$\bolds{\rho}\in\mathbb{P}(\Omega_M)$, here we consider the performance
gap specifically at the uniform vector in the information state space.
%The reason is that $\Omega_M$
%%and hence, the information state space $\mathbb{P}(\Omega_M)$
%grows as $M$ increases and a belief state $\bolds{\rho}\in
}
\[
\lim_{{L},M \to\infty} \frac{V_{\pi}(\bolds{\rho}_{u,M}) -
V^*(\bolds{\rho}_{u,M})}{V_{\pi}(\bolds{\rho}_{u,M})}<1,\qquad \lim
_{{L},M \to\infty} \frac{V_{\pi}(\bolds{\rho}_{u,M}) -
V^*(\bolds{\rho}_{u,M})}{V_{\pi}(\bolds{\rho}_{u,M})}=0.
\]
\end{definition}

%Under Assumptions~\ref{KL0} and \ref{JumpM},
%for $\L> \frac{\log M}{I_{\max}(M)}$, and if $
%policy $\tilde{\pi}_2$ is order optimal in $\L$ and $M$.
%%
%The proof simply follows from Definition~\ref{AsymM}, Corollary~

%Note that a sufficient condition for $\underline{I}_2>0$ can be
%obtained by strengthening
%Assumption 1 in the following manner: there exists $\zeta>0$ such that
%for any
%hypothesis $H_i$, there exists an action $a\in\mathcal{A}_M$ for
%which $\min_{q\in Q^a_{-i}} D(q_i^a \| q)\ge\zeta$ where
%for all $i \in\Omega_M, a \in\mathcal{A}_M$, $Q^a_{-i}$ is
%the convex hull of distributions $\{q^a_j(\cdot)\}_{j\in\Omega_M- \{i

%
%co4 #&#
\begin{corollary}
\label{corAsym2LM} Under Assumptions~\ref{KL0}, \ref{Jump} and
\ref{JumpM}, for ${L}> \frac{\log M}{I_{\max}(M)}$, and if
\mbox{$\underline{I}_2>0$}, policy\vspace*{1pt} $\tilde{\pi}_2$ is order optimal in ${L}$
and $M$.
%Under Assumptions~\ref{KL0} and \ref{JumpM}, for $\L> \frac{\log
%M}{I_{\max}(M)}$, and
Furthermore, if $\widebar{I}_{\max} = \underline{I}_2$
and $\widebar{D}_{\max} = \underline{D}_{2}$, %and $
policy $\tilde{\pi}_2$ is asymptotically optimal of order-1 in ${L}$
and $M$.
\end{corollary}

\begin{pf}
The proof follows from Definition~\ref{AsymM}, Corollary~\ref{CorLB2}
and Theorem~\ref{thmUB2}.
\end{pf}

%asymptotic results could be obtained
%if $\sup_M$ and $\inf_M$ are replaced by $\limsup_{M\to\infty}$ and $
%More precisely, under the above modifications, the dominating terms
%(in $\L$ and $M$) in our lower and upper bounds remain unchanged.
%}

%%%%%%%%%%%%%%%%%%%%%%%%%%%%%%%%%%%%%%%%%%%%%
%Information Acquisition Rate and Reliability
%%%%%%%%%%%%%%%%%%%%%%%%%%%%%%%%%%%%%%%%%%%%%
%s5.3 #&#
\subsection{Information acquisition rate and reliability}\label{Rate}

%In this section, we go back to Problem~(P') and use the obtained bounds
%to extend the notions of achievable (communication) rate and error
%exponent in the context of sequential hypothesis testing.

In this section we explain the primal (constrained) version of
Problem~\ref{probp}, referred to as Problem~\ref{probp'}, and use the
obtained bounds in Section~\ref{SecB} to extend the (information
theoretic) notions of achievable communication rate and error exponent
to the context of active sequential hypothesis testing.\vadjust{\goodbreak}
%in Appendix~\ref{AppCode}.
%

{\renewcommand{\theprob}{(P$'$)}
\begin{prob}[(Information acquisition problem)]\label{probp'}
Consider a sequence of active hypothesis testing problems
    indexed by parameter $M$ (i.e., the number of hypotheses of interest), action space
    $\mathcal{A}_M$ and observation kernels
    $\{q^a_i(\cdot)\}_{i\in\Omega_M,a\in \mathcal{A}_M}$: a
    Bayesian decision maker with uniform prior belief $\bolds{\rho
    }(0)=\bolds{\rho}_{u,M}$ is responsible to find the true
    hypothesis with the objective to
%e17 #&#
\begin{equation}
\label{Obj}
\mbox{minimize } \mathbb{E} [ \tau ] \mbox{ subject to }
\operatorname{Pe}\le\varepsilon,
\end{equation}
where $\tau$ is the stopping time at which the decision maker
retires, $\operatorname{Pe}$ is the probability of making a wrong
declaration, and $\varepsilon>0$ denotes the desired probability of
error. Furthermore, let the set of actions and observation kernels
grow monotonically as $M$ increases, that is, for all $M<M'$,
%e18 #&#
\begin{equation}
\label{MonotonicGrowth2} \mathcal{A}_M\subseteq\mathcal{A}_{M'}\quad
\mbox{and}\quad \bigl\{ q^a_i(\cdot)\bigr\}_{i\in\Omega_M,a\in\mathcal{A}_M}
\subseteq\bigl\{ q^a_i(\cdot)\bigr\}_{i\in\Omega_{M'},a\in\mathcal{A}_{M'}}.
\end{equation}
\end{prob}}

%Consider a policy $\pi$.
Let $\mathbb{E}_{\pi}[\tau]$ and $\operatorname{Pe}_{\pi}$ denote,
respectively, the expected stopping time (or, equivalently, the
expected number of collected samples) and the probability of error
under policy $\pi$. Following the notation in~\cite{Polyanskiy11}, we
define $M_{\pi }(t,\varepsilon)$ as the maximum number of hypotheses
among which policy $\pi$ can find the true hypothesis with
$\mathbb{E}_{\pi}[\tau] \le t$ and $\operatorname{Pe}_{\pi}
\le\varepsilon$. Policy $\pi$ is said to achieve information
acquisition rate $R>0$ with reliability (also known as error exponent)
$E>0$ if
%
%e19 #&#
\begin{equation}
\lim_{t\to\infty} \frac{1}{t} \log M_{\pi}
\bigl(t,2^{-Et}\bigr) = R.
\end{equation}
For a fixed number of hypotheses $M$, hence at information acquisition
rate $R=0$, policy $\pi$ is said to achieve reliability $E>0$ if
%
%e20 #&#
\begin{equation}
\lim_{t\to\infty} \frac{-1}{t} \log\operatorname{Pe}_{\pi}(t,M)
= E,
\end{equation}
where $\operatorname{Pe}_{\pi}(t,M)$ is the minimum probability of
error that policy $\pi$ can guarantee for $M$ hypotheses with the
constraint $\mathbb{E}_{\pi}[\tau] \le t$.

The reliability function $E(R)$ is defined as the maximum achievable
error exponent at information acquisition rate~$R$.

%%%%%%%%%%%%%%%%%%
%%%%%%%%%%%%%%%%%%
%
%f1 #&#
\begin{figure}[b]

\includegraphics{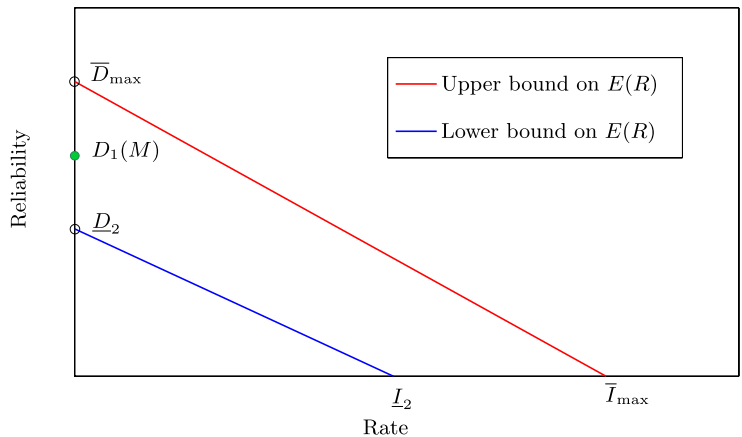}

%%\psfrag{R}{Rate}
%%\psfrag{E}{Reliability}
%%\psfrag{Im}{$\widebar{I}_{\max}$}
%%\psfrag{Dm}{$\widebar{D}_{\max}$}
%%\psfrag{I2}{$\underline{I}_2$}
%%\psfrag{D1}{$D_1(M)$}
%%\psfrag{D2}{$\underline{D}_2$}
%%\psfrag{LLLLLLLLLL}{Lower bound on $E(R)$} %{$\tilde{D}_2 \left(1 -
%%\psfrag{UUUUUUUUUU}{Upper bound on $E(R)$} %{$D_{\max}
\caption{Lower and upper bounds on the optimal reliability function~$E(R)$.}\label{figER}
\end{figure}

%%%\ignore{ Next we use the bounds obtained in Section~\ref{SecB} as well
%%%as Lemma~\ref{CodeLemmaLB} to characterize}

%Before we proceed, we refer the reader to Tables~\ref{ListNote1} and~
Before we proceed with the upper and lower bounds on the maximum
achievable information acquisition rate and the optimal reliability
function, we refer the reader to Table~\ref{ListNote1} for the list of
notation introduced in Section~\ref{SecB}.
Also recall that $D_1(M)$ and $D_2(M)$ denote, respectively, the
harmonic mean of $\{ D_{\bolds{\mu}_i}(M) \}_{i\in\Omega_M}$ and $\{
D_{\bolds{\eta}_i}(M) \}_{i\in\Omega_M}$.

%Also let $D_1(M)$ and $D_2(M)$ denote respectively the harmonic mean of
%$\{ D_{\bolds{\mu}_i}(M) \}_{i\in\Omega_M}$ and
%$\{ D_{\bolds{\eta}_i}(M) \}_{i\in\Omega_M}$, i.e.,
%%
% D_1(M) = M \left(\sum_{i=1}^M \frac{1}{D_{\bolds{
% D_2(M) = M \left(\sum_{i=1}^M \frac{1}{D_{\bolds{
%%
%%Let $\underline{D}_1$ and $\underline{D}_2$ denote respectively the
%limits of $D_1$ and $D_2$ as $M\to\infty$.
%Moreover, let
%%
%% \underline{D}_1 = \inf_{M} D_1, \quad
% \underline{D}_2 = \inf_{M} D_2(M).

% \tilde{D}_1 = \frac{M}{\sum_{i=1}^M \frac{1}{\max_{
% \tilde{D}_2 = \frac{M}{\sum_{i=1}^M \frac{1}{\max_{
%
%Also let
%%
%%\begin{eqnarray*}
% $\tilde{I}_2 = \max_{\bolds{\lambda} \in\mathbb{P}(
%%\end{eqnarray*}
%%
%Recall that $I_{\max}={\max_{a \in\mathcal{A}_M} \max
%_{\hat{\bolds{\rho}}\in\mathbb{P}(\Omega_M)}
%I(\hat{\bolds{\rho}};q^{a}_{\hat{\bolds{\rho}}})}$ and
%$D_{\max}={\max_{i,j\in\Omega_M} \max_{a \in

%Suppose the hypotheses are equiprobable, i.e., $\rho_i(0)=
%%$P(\{\theta=i\})=\frac{1}{M}$, $\forall i \in\Omega_M$.
%No policy can achieve positive reliability $E>0$ at rates higher than $
%acquisition rate,
%E(R) \le\begin{cases}
%D_1(M) & \mbox{fixed $M$ $(R=0)$}

%
%co5 #&#
\begin{corollary}
\label{converse}
%Suppose the hypotheses are equiprobable, i.e., $\rho_i(0)=
For any given fixed $M$ (rate $R=0$), no policy can achieve reliability
higher than $D_1(M)$. Also, no policy can achieve positive reliability
\mbox{$E>0$} at rates higher than $\widebar{I}_{\max}$. Furthermore,
%e21 #&#
\begin{equation}
E(R) \le\widebar{D}_{\max} \biggl(1-\frac{R}{\widebar{I}_{\max}} \biggr),\qquad R\in(0,
\widebar{I}_{\max}).
\end{equation}
\end{corollary}

%
%re4 #&#
\begin{remarks}
Corollary~\ref{converse} establishes an upper bound,
$\widebar{I}_{\max}$, on the maximum achievable information acquisition
rate. As shown in the supplemental article~\cite{Naghshvar-SuppA},
Section~3.3, this result can be strengthened to show
that no policy can achieve diminishing error probability at rates
higher than $\widebar{I}_{\max}$.
%The sketch of the proof is provided in Appendix~\ref{AppCode}.
\end{remarks}

%
%co6 #&#
\begin{corollary}
\label{D1} For fixed $M$, hence at rate $R=0$, a policy $\pi$ can
achieve the maximum reliability, that is, $E=D_1(M)$, if and only if it
is asymptotically optimal (of order-1 or higher) in $L$. Furthermore, a
policy $\pi$ can achieve a nonzero rate $R>0$ with nonzero reliability
$E>0$
only if it is order optimal in ${L}$ and $M$. %solution to Problem~(P).
\end{corollary}

%A policy $\pi$ can achieve a non-zero rate $R>0$ with non-zero
%reliability $E>0$
%if and only if it is order optimal in $\L$ and $M$. %solution to
%Problem~(P).
%
%Corollaries~\ref{D1} and~\ref{iffordopt} imply that:
%1) for fixed $M$, hence at $R=0$, policies $\tilde{\pi}_1$ and $\pi^*$
%achieve the optimal error exponent; and
%2) policies $\tilde{\pi}_2$ and $\pi^*$ achieve strictly positive
%information acquisition rate with exponential decaying probability of
%error.
%In particular,

Corollary~\ref{D1} implies that for fixed $M$, hence at $R=0$, policies
$\tilde{\pi}_1$ and $\pi^*$ achieve the optimal error exponent, while
policy $\tilde{\pi}_2$ might or might not [depending on
condition~(\ref{pi2asym01})]. Furthermore, Corollary~\ref{D1}, in
effect, underlines the deficiency of characterizing the solution to
Problems~\ref{probp} in terms of $L$ in isolation from $M$, hence,
Chernoff's notion of asymptotic optimality (solely in $L$). In
particular, an order optimal policy can achieve nonzero rate and
reliability simultaneously, an improvement over $\tilde{\pi}_1$ (and
all extensions of~\cite{Chernoff59}).\looseness=-1

%
%co7 #&#
\begin{corollary}
\label{I2D2}
%Suppose the hypotheses are equiprobable, i.e., $\rho_i(0)=
%$P(\{\theta=i\})=\frac{1}{M}$, $\forall i \in\Omega_M$.
Policy $\tilde{\pi}_2$ achieves rate $R\in[0,\underline{I}_2]$ with
reliability $E$ if
%e22 #&#
\begin{equation}
\label{E2reliab} E \le\underline{D}_2 \biggl(1-\frac{R}{\underline{I}_2}
\biggr).
\end{equation}
\end{corollary}

Figure~\ref{figER} summarizes the results above. The upper bound on the
reliability function is shown in red. Policy $\tilde{\pi}_1$ achieves
the optimal reliability $D_1(M)$ for fixed $M$ (at $R=0$) with no
provable guarantee for $R>0$ (this point is shown in green), while
policy $\tilde{\pi}_2$ ensures an exponentially decaying error
probability (the error exponent is shown in blue) for
$R\in[0,\underline{I}_2)$.\vadjust{\goodbreak}

%
%re5 #&#
\begin{remarks}
It can be shown that any optimal policy $\pi^*$ for Problem~\ref{probp}
also achieves any rate $R\in[0,\underline{I}_2]$ with reliability $E$
satisfying~(\ref{E2reliab}) for Problem~\ref{probp'}.
%The proof is provided in Appendix~\ref{AppCode}.
\end{remarks}

The proofs of all the results in this section are provided in the
supplemental article~\cite{Naghshvar-SuppA}, Section~3,
and are based on the fact that Problem~\ref{probp} can be viewed as a
Lagrangian relaxation of Problem~\ref{probp'}. It is somewhat intuitive
that as ${L}\to\infty$ the solution of Problem~\ref{probp} is closely
related to that of Problem~\ref{probp'} when $\varepsilon\to0$. The
following lemma makes this intuition precise.

%
%le2 #&#
\begin{lemma}
\label{CodeLemmaLB}
%Consider Problem~(P') described in Section~\ref{sum}.
Let $\mathbb{E} [\tau^*_{\varepsilon}]$ denote the minimum expected
number of samples required to achieve
$\operatorname{Pe}\le\varepsilon$. We have
%
%e23 #&#
\begin{equation}
\mathbb{E} \bigl[\tau^*_{\varepsilon}\bigr] \ge(1 - \varepsilon{L}) \bigl(V^*
\bigl(\bolds{\rho}(0)\bigr)-1\bigr),
\end{equation}
%
%where $\underline{V_2}(\bolds{\rho})$ is the lower bound
%(Propositions~\ref{thmLB2})
%associated with the solution to Problem~(P) for prior belief $
where $V^*(\bolds{\rho}(0))$ is the optimal solution to
Problem~\ref{probp} for prior belief $\bolds{\rho}(0)$ and penalty of
wrong declaration~${L}$.
\end{lemma}

Given the above connection, Corollary~\ref{converse} follows readily
from the lower bounds obtained in Proposition~\ref{propLB1} and
Theorem~\ref{thmLB2} (in particular, its Corollary~\ref{CorLB2}), and
Corollaries~\ref{D1} and~\ref{I2D2} follow from the upper bounds given
by Proposition~\ref{propUB1} and Theorem~\ref{thmUB2}.
\section{Examples}
\label{SecExmp}

In this section we consider important special cases of the active
hypothesis testing to provide some intuition about the conditions of
Corollaries \ref{corAsym1}~and~\ref{corAsym2LM}, and, in particular,
establish the order-2 asymptotic optimality of $\tilde{\pi}_2$ for a
fixed value of $M$ and rate--reliability optimality of policy
$\tilde{\pi}_2$.
%and establish the asymptotic optimality of the proposed heuristics.
%for which the upper and lower bounds are asymptotically tight,
%establishing the asymptotic optimality of the proposed heuristics.
%By asymptotically tight, we mean bounds whose gap grows
%sub-logarithmically as $\L$ or $M$ increases (or equivalently with
%increasing number of samples), and hence, the gap grows in a much
%slower rate than the bounds themselves.

%%%%%%%%%%%%%%%%%%%%%%%%%%%%%%%%%%%%%%%%%%%%%%%%%%%%%%%%%%%%%%%%%%%%%%%%%%%%%%%%%%%
%%% Binary Hypothesis Testing
%%%%%%%%%%%%%%%%%%%%%%%%%%%%%%%%%%%%%%%%%%%%%%%%%%%%%%%%%%%%%%%%%%%%%%%%%%%%%%%%%%%
%s6.1 #&#
\subsection{Binary hypothesis testing} %($M=2$)

Consider Problem~\ref{probp} for $M=2$. In this setting, %as noted in Remark~
policies $\tilde{\pi}_1$ and $\tilde{\pi}_2$ are equivalent and by
Corollary~\ref{corAsym1}, both policies are asymptotically optimal of
order-1 in ${L}$.
Asymptotic optimality of order-2 of $\tilde{\pi}_1$ and $\tilde{\pi
}_2$ is also verified from Corollary~\ref{corAsym1} %Theorem~
since equality~(\ref{pi2asym01}) holds trivially for $M=2$.
Furthermore, we obtain %the bounds are asymptotically tight (the gap
%between the bounds remains fixed as $\L$ grows) and we have,
\[
V^*(\bolds{\rho})= \rho_1 \frac{\log{L}- \log (\rho_1/\rho_2)}{\max_{a \in\mathcal{A}_M} D(q^{a}_1 \| q^{a}_2)} +
\rho_2 \frac{\log{L}- \log(\rho_2/\rho_1)}{\max_{a \in\mathcal{A}_M} D(q^{a}_2 \| q^{a}_1)} + O(1).
\]
%
% $$V^*(\bolds{\rho})=
% \rho_1 \frac{\log\frac{1-\L^{-1}}{\L^{-1}} - \log\frac{\rho_1}{
% + \rho_2 \frac{\log\frac{1-\L^{-1}}{\L^{-1}} - \log\frac{\rho_2}{
%O(1).$$
%
% This is the case that we studied in \cite{Allerton10}.
% %In \cite{ISIT10}, we provided a sufficient condition under which
%active binary hypothesis testing reduces to a passive test where one
%action dominates the other ones.
% In \cite{Allerton10}, we proposed a policy (similar to $\tilde{\pi}$
%defined in Subsection~\ref{SecUB}) whose total cost was shown to be
%only a constant off from the optimal total cost.
% There have been extensive studies on active binary hypothesis testing
%from a non-Bayesian point of view.
% In a recent paper \cite{PolyanskiyITA2011} by Polianskey et al.,
%authors studied the optimal tradeoff between the achievable exponents
%of two types of errors for this problem.
%

%This %together with characterization of $\widebar{V}_1$,
%establishes asymptotic optimality of $\tilde{\pi}_1$ (the same of $
%
The problem of reliability (error exponent) associated with passive
binary hypothesis testing
with fixed-length (nonsequential) as well as variable-length
\mbox{(sequential)} sample size has been studied
by~\cite{Blahut74,Csiszar04,Haroutunian07}. %NitinawaratArxiv}.
The generalization to channel coding with feedback with two messages
was addressed in~\cite{Berlekamp64,Nakiboglu12,Berlin09}. Recently, the
authors in~\cite{Hayashi09} and \cite{PolyanskiyITA2011} have
generalized this problem for fixed-length and variable-length
sample
size, respectively, to the active binary hypothesis testing in the
non-Bayesian context, and identified the error exponent\vadjust{\goodbreak} corresponding
to both error types.
% and shown the optimality of policy $\tilde{\pi}_1$ (in terms of error
%exponent).
%Our work complements the findings in~\cite{PolyanskiyITA2011} by
%providing
%an asymptotic optimal solution in a total cost (and Bayesian) sense.
%the asymptotic optimality of $\tilde{\pi}_1$ in a total cost (and
%Bayesian) sense.
Our work provides nonasymptotic bounds as well as an asymptotic optimal
solution in a total cost and Bayesian sense, and is consistent with the
findings in~\cite{PolyanskiyITA2011}.

%%%%%%%%%%%%%%%%%%%%%%%%%%%%%%%%%%%%%%%%%%%%%%%%%%%%%%%%%%%%%%%%%%%%%%%%%%%%%%%%%%%
%%% Noisy Dynamic Search
%%%%%%%%%%%%%%%%%%%%%%%%%%%%%%%%%%%%%%%%%%%%%%%%%%%%%%%%%%%%%%%%%%%%%%%%%%%%%%%%%%%
%s6.2 #&#
\subsection{Noisy dynamic search} %(fixed $M$, asymptotic in $\L$)}
%%($M<\infty$)}
\label{secNDS} Consider the problem of sequentially searching for a
single target in $M$ locations where the goal is to find the target
quickly and accurately. In each step, the player can inspect an
allowable combination of the locations, and the outcome of the
inspection is noisy.
% The collection of allowable combinations of the locations that can be
%visually inspected
%in one time slot is specified by the \emph{inspection space}.
% In the special case that the inspection space is composed of only
%single segments,
This problem is closely related to the problems of \textit{fault
detection}, \textit{whereabouts search} and \textit{group testing}. In
fault detection, the objective is to determine the faulty component in
a system known to have one failed component \cite{Nachlas90,Castanon95}. % where each component can be tested
%independently.
In whereabouts search, the goal is to find an object which is hidden in
one of $M$ boxes, where it is usually assumed that there is no
\textit{false alarm}, that is, the outcome of inspecting box $i$ is
always $0$ if no object is present, and is a Bernoulli random variable
with a
known parameter otherwise \cite{Tognetti68,Kadane71}. % where only
%one box can be searched at a time.
In group testing, the goal is to locate the nonzero
element\footnote{Group testing with $d>1$ nonzero elements is also a special
case of active hypothesis testing with~${M\choose d}$ hypotheses
(possible configurations).} of a vector in $\mathbb{R}^M$
%(i.e., to find the true configuration out of $M = N \choose d$
%possible ones)
with a possible noisy linear measurement of the vector
\cite{Sejdinovic10,Chan11}.
%
%One possible search strategy for these problems is the maximum
%likelihood policy which inspects a
%segment with the highest probability of having the hidden target.
One possible search strategy for these problems is the maximum
likelihood policy.
In the case of fault detection/whereabouts search, this policy is
equivalent to one that inspects a
segment with the highest probability of having the faulty
component/hidden object,
while in the case of group testing, it is equivalent to measuring the
most likely nonzero element of the vector.
%using a measurement vector with $d$ non-zero elements matching the
%most likely configuration.
However, as the number of segments or the dimension of vectors, $M$,
increases, the scheme becomes impractical. In such a case, it is more
intuitive to initially follow a noisy binary search
\cite{Horstein63,Burnashev74,Nowak11IT} and narrow down the search to
single segments only after we have collected sufficient information
supporting the presence of the target in those
segments~\mbox{\cite{Posner63,Stone75}}.

In this section, we first consider the problem of a noisy dynamic
search with size-dependent Bernoulli noise whose special cases have
been independently studied in
\cite{Horstein63,Tognetti68,Kadane71,Burnashev74,Nachlas90,Sejdinovic10,Chan11,Nowak11IT}.\footnote
{Of course in this paper we are interested in a sequential setting
where the sample size is not fixed a priori and is determined by the
observation outcomes.}
% whose special cases include the problems mentioned above and have
%been studied extensively in \cite{}.
Remark~\ref{ByndBern} at the end of this section discusses a
generalization for the symmetric noise model of \cite{Castanon95}.
%at the end of this section explains how the results can be extended
%beyond the Bernoulli noise model so long as the observation outcomes
%are symmetric as in \cite{Castanon95}.

Let $a \subset\Omega_M$ be a subset of locations that can be
simultaneously inspected, referred to as the inspection region
hereafter, and let $\mathcal{A}_M= 2^{\Omega_M}$ be the collection of
all allowable inspection regions. We assume that the outcome of an
inspection depends on the size of the inspection region. More
precisely, the outcome of inspecting region $a$, where $|a|=n$, is a
random variable with Bernoulli distribution:
\begin{eqnarray*}
q^{a}_i= \cases{ \mathcal{B}(1-p_n), &\quad
if $i\in a$ and $|a|=n$
\vspace*{2pt}\cr
\mathcal{B}(p_n), &\quad if $i\notin a$
and $|a|=n$}\qquad\forall i\in\Omega_M, \forall a\in
\mathcal{A}_M,
\end{eqnarray*}
where $p_1 > 0 $ and for all $n$, $p_n \le p_{n+1}$ and $p_n \leq p$
for some $p<0.5$.\vadjust{\goodbreak}
%

%
%le3 #&#
\begin{lemma}
\label{LemmaNDS}
%Consider the problem of noisy dynamic search in the case of Bernoulli
%noise where $f_n=\mathcal{B}(1-p_n)$, $\bar{f_n}=\mathcal{B}(p_n)$, $0
%< p_n \le p_{n+1}$, and $p_{\infty}:=\sup_n p_n < 0.5$.
Consider the problem of a noisy dynamic search with size-dependent
Bernoulli noise explained above. We have
%for all $i\in\Omega_M$, %$\forall i\in\Omega_M$
%e24 #&#
%e25 #&#
%e26 #&#
\begin{eqnarray}
\qquad && \min_{j \neq i} \max_{a \in\mathcal{A}_M} D
\bigl(q^{a}_i\|q^{a}_j\bigr) =
{D}_{\bolds{\eta}_i}(M) = (1-2p_1)\log\frac{1-p_1}{p_1}\qquad \forall i\in\Omega_M,\label{NDS01}
\\
\label{NDS02} && \underline{D}_{2} = \widebar{D}_{\max} =
(1-2p_1)\log \frac{1-p_1}{p_1},
\\
\label{NDS03}  && 0 < 1- \sup_n H\bigl([p_n,1-p_n]
\bigr) \le\underline{I}_2 \le\widebar{I}_{\max} \le1- H
\bigl([p_1,1-p_1]\bigr).
\end{eqnarray}
\end{lemma}

%The proof of Lemma~\ref{LemmaNDS} is provided in the supplemental
The proof is provided in the supplemental
article~\cite{Naghshvar-SuppA}, Section~4.

Lemma~\ref{LemmaNDS}, together with Corollaries~\ref{corAsym1}
and~\ref{corAsym2LM}, implies that $\tilde{\pi}_2$ attains
asymptotic optimality of order-2 in ${L}$, and order optimality
in ${L}$ and $M$. Furthermore, for the special case of size-independent
Bernoulli noise where $0<p_1=p_2=\cdots=p<0.5$, policy $\tilde{\pi}_2$
attains asymptotic optimality of order-1 in ${L}$ and~$M$.

%Under the condition of Lemma~\ref{LemmaNDS},
The active hypothesis testing scheme proposed by
Chernoff~\cite{Chernoff59} as well as its
variants~\cite{Bessler60,Blot73}, when specialized to a noisy dynamic
search with size-independent Bernoulli noise, simplifies to one that
inspects, at each instant, a location with the highest probability of
having the target.
%The schemes proposed in~\cite{Chernoff59,Blot73,NitinawaratArxiv}
%(and policy $\tilde{\pi}_1$),
%when specialized to noisy dynamic search with Bernoulli noise,
%simplifies to one that inspects, at each instant (in the second phase),
%a location with the highest probability of having the target.
This scheme, which was also studied in~\cite{Castanon95} in a finite
horizon context,
%with symmetric observations and was shown to
%be optimal among policies that can only inspect a single location at a
%time,
has an information acquisition rate that is restricted to zero, while
at zero rate, it achieves asymptotic optimality and maximum error
exponent $\widebar{D}_{\max}=(1-2p)\log\frac {1-p}{p}$. In contrast,
in~\cite{Horstein63,Burnashev74}, a noisy binary search was proposed in
which the locations are partitioned along the median of the posterior
and, in effect, are inspected along a generalized binary tree. It was
shown in~\cite{Horstein63,Burnashev74} that
% in the special case of size-independent Bernoulli noise where for all
%$n$, $p_n=p$, $p\in(0,0.5)$,
the proposed policy can achieve any rate
$R < 1- H([p,1-p])$ with reliability
%$R < \tj{D(f \| \frac{1}{2}f + \frac{1}{2}\bar{f})=?}$ with reliability
$E(R) = 1- H([p,1-p]) - R$. In other words, the proposed policy
in~\cite{Horstein63,Burnashev74} is asymptotically optimal in $M$
(since $1- H([p,1-p]) = \widebar{I}_{\max}$) but only order optimal in
${L}$ (since $ 0 < 1- H([p,1-p]) < (1-2p)\log\frac{1-p}{p} =
\widebar{D}_{\max}$).
Lemma~\ref{LemmaNDS} shows that, in the case of size-independent
Bernoulli noise, our proposed policy $\tilde{\pi}_2$ combines the best
of the above two approaches: in its first phase, by randomly selecting
actions from $\mathcal {A}_M$, it ensures the maximum acquisition rate
obtained by the noisy binary search of~\cite{Horstein63,Burnashev74},
while its second phase coincides with the schemes
in~\cite{Chernoff59,Bessler60,Blot73}, ensuring the maximum feasible
error exponent.

%hence ensuring the maximum acquisition rate and the maximum feasible
%error exponent simultaneously.
%%in its first phase, by randomly selecting actions from $
%%maximum acquisition rate $1- H([p,1-p])$ obtained by the generalized
%binary
%%search of~\cite{Horstein63,Burnashev74}; while its second phase
%coincides with the schemes in~\cite{Chernoff59,Blot73,%NitinawaratArxiv} ensuring the maximum feasible error exponent $(1-2p)

%
%re6 #&#
\begin{remarks}
\label{ByndBern} Lemma~\ref{LemmaNDS} can be extended beyond the
Bernoulli noise model so long as the observation kernels
\begin{eqnarray*}
q^{a}_i(\cdot)= \cases{ f_{n}(\cdot), &\quad
if $i\in a$ and $|a|=n$
\vspace*{2pt}\cr
\bar{f}_{n}(\cdot), &\quad if $i\notin a$
and $|a|=n$}\qquad\forall i\in\Omega_M, \forall a\in
\mathcal{A}_M,
\end{eqnarray*}
satisfy the following conditions:
%e27 #&#
%e28 #&#
%e29 #&#
\begin{eqnarray}
\label{asmsym} %\mbox{Symmetry: }
&& f_n(z) = \bar{f}_n(b-z)\qquad \forall z\in\mathcal{Z}\mbox{ for some } b\in\mathbb{R},
\\
&& D(f_n\|\alpha f_n + \bar{
\alpha} \bar{f}_n) \ge D(f_{n+1}\|\alpha f_{n+1}
+ \bar{\alpha} \bar{f}_{n+1})\qquad \forall\alpha\in[0,1], \bar{\alpha}=1-\alpha,\hspace*{-29pt} \label{asmmon}
\\
\label{asmjumps} %\mbox{Bounded jumps: }
&& \sup_n \sup_{z\in\mathcal{Z}} f_{n}(z)/\bar{f}_{n}(z) <\infty. %
\end{eqnarray}
In particular, under these conditions %we obtain, %and for all $i\in
%e30 #&#
%e31 #&#
%e32 #&#
\begin{eqnarray}
\label{NDSgen01}  && \min_{j \neq i} \max_{a \in\mathcal{A}_M} D
\bigl(q^{a}_i\|q^{a}_j\bigr) ={D}_{\bolds{\eta}_i}(M) = D(f_1 \| \bar{f}_1)\qquad\forall i\in\Omega_M,
\\
\label{NDSgen02} && \underline{D}_{2} = \widebar{D}_{\max} =
D(f_1 \| \bar {f}_1),
\\
\label{NDSgen03}  && \inf_n D\biggl(f_n \bigg\|
\frac{1}{2}f_n + \frac{1}{2}\bar{f}_n
\biggr) \le \underline{I}_2 \le\widebar{I}_{\max} \le D
\biggl(f_1 \bigg\| \frac
{1}{2}f_1 + \frac{1}{2}
\bar{f}_1\biggr).
\end{eqnarray}

Condition~(\ref{asmsym}) implies that given a fixed inspection area,
the collected samples provide identical information regarding the
presence of the target or its absence. %; and
%This condition is satisfied if, for instance, observation outcomes are
%modeled as a signal plus
%noise, where the signal component appears only if the target is
%included in the inspection region,
%and the noise distribution is symmetric with respect to its mean value.
Condition~(\ref{asmmon}) implies that the samples become less
informative as the size of the inspection region increases. %; and
Conditions~(\ref{asmsym}) and~(\ref{asmmon}) are natural, while
condition~(\ref{asmjumps}) is a technical one to ensure that
Assumptions~\ref{Jump} and~\ref{JumpM} hold (we address weakening
these assumptions in Section~\ref{WeakenAsmp}).
%condition~\eqref{asmjumps} restricts the amount of information
%corresponding to each inspection region, and ensures that Assumptions~
%assumptions in Section~\ref{WeakenAsmp}).
\end{remarks}

%%%%%%%%%%%%%%%%%%%%%%%%%%%%%%%%%%%%%%%%%%%%%%%%%%%%%%%%%%%%%%%%%%%%%%%%%%%%%%%%%%%
%%% Weakening Assumption 2
%%%%%%%%%%%%%%%%%%%%%%%%%%%%%%%%%%%%%%%%%%%%%%%%%%%%%%%%%%%%%%%%%%%%%%%%%%%%%%%%%%%
%s7 #&#
\section{Discussions}
\label{WeakenAsmp}

In this section we provide a discussion on the technical assumptions of
the paper. In particular, we discuss the necessity of our
Assumptions~\ref{KL0} and~\ref{Jump}, and compare them with the common
assumptions in the literature. In contrast to Assumption~\ref{KL0}
which is shown to be necessary for the problem of active hypothesis
testing to have a meaningful solution, Assumption~\ref{Jump} can be
relaxed to more general assumptions without affecting the asymptotic
results of the paper.

%s7.1 #&#
\subsection{\texorpdfstring{Assumption~\protect\ref{KL0}}{Assumption 1}}
We first discuss the necessity of Assumption~\ref{KL0}. If
Assumption~\ref{KL0} does not hold, then there exist two hypotheses
$i,j\in\Omega_M$, $i\neq j$ such that for all $a\in\mathcal{A}_M$,
$D(q^a_i\|q^a_j)=0$. In other words, $q^a_i(\cdot)=q^a_j(\cdot)$ for
all $a\in\mathcal {A}_M$, and, hence, the decision maker is not capable
of distinguishing these two hypotheses. In this sense,
Assumption~\ref{KL0} is necessary for Problem~\ref{probp} to be
meaningful.

Next we compare Assumption~\ref{KL0} to its counterpart
in~\cite{Chernoff59}:
%which requires that
%$D(q_i^a\|q_j^a)>0$, $\forall i,j\in\Omega_M$, $i\neq j$, $\forall a\in
%In other words, any sensing action $a\in\mathcal{A}_M$ in~
%pairs.

{\renewcommand{\theass}{1$'$}
\begin{ass}\label{ass1'}
$D(q_i^a\|q_j^a)>0$, $\forall i,j\in\Omega _M$, $i\neq j$, $\forall
a\in\mathcal{A}_M$.
\end{ass}}

This assumption assures \textit{consistency} (see Lemma~1 in \cite{Chernoff59}),
that is,\break $\argmax_{i\in\Omega_M} \rho_i(t)$ converges
exponentially fast to the true hypothesis regardless of the way the
sensing actions are selected. However, this assumption is very
restrictive and does not hold in many problems of interest such as
channel coding with feedback\vadjust{\goodbreak} \cite{Burnashev76} and noisy dynamic
search (e.g., one cannot discriminate between locations 1 and 2 by
inspecting location 3). It was remarked in \cite{Chernoff59},
Section~7, that the above restrictive assumption can be relaxed if the
proposed scheme is modified to take a (possibly randomized) action
capable of discriminating between all hypotheses pairs infinitely often
(e.g., at any time $t$ when $t$ is a perfect square).
%Next we compare Assumption~\ref{KL0} to its counterpart in~
%$D(q_i^a\|q_j^a)>0$, $\forall i,j\in\Omega_M$, $i\neq j$, $\forall a\in
%In other words, any sensing action $a\in\mathcal{A}_M$ in~
%pairs which is very restrictive and does not hold in many problems of
%interest such as channel coding with feedback and noisy dynamic search
%(for example one cannot discriminate between locations 1 and 2 by
%inspecting location 3).
%It was remarked in \cite{Chernoff59}, Section~7, that the above
%restrictive assumption can be relaxed if the proposed scheme is
%modified to take a (possibly randomized) action capable of
%discriminating between all hypotheses pairs infinitely often (for
%example at any time $t$ when $t$ is a perfect square).
%However, here we have shown that such a condition is not necessary by
%constructing
%policy $\tilde{\pi}_1$, which is a simple two-phase modification of
%Chernoff's original scheme in which
In this paper, however, we took a different approach and constructed
policy $\tilde{\pi}_1$, a~simple two-phase modification of Chernoff's
original scheme in which testing for the maximum likely hypothesis is
delayed and contingent on obtaining a certain level of confidence.
%%More specifically, in its first phase, $\tilde{\pi}_1$ selects a
%(possibly randomized) action under which all pairs of hypotheses can
%be distinguished from each other; while its second phase coincides
%with Chernoff's original scheme~\cite{Chernoff59} where only the pairs
%including the most likely hypothesis are considered.

%In this sense, the main contribution of the above facts, hence policy $
%relax the technical assumption in~\cite{Chernoff59} which requires
%$D(q_i^a\|q_j^a)>0$, $\forall i,j\in\Omega_M$, $i\neq j$, $\forall a\in
%and replace it with Assumption~\ref{KL0}.

%s7.2 #&#
\subsection{\texorpdfstring{Assumption~\protect\ref{Jump}}{Assumption 2}}
We first discuss the necessity of Assumption~\ref{Jump}.
%In this paper, our focus has been on the problem of active hypothesis
%testing with noisy observations.
For observation kernels with bounded support, Assumption~\ref{Jump} is
a necessary condition to ensure that the observation kernels are
absolutely continuous with respect to each other and, hence, no
observation is noise free. Although this assumption might hold in many
settings such as the problem of a noisy dynamic search with Bernoulli
noise explained in Section~\ref{secNDS}, it does not hold in general
for observation kernels with unbounded support such as Gaussian
distribution.
%
%In this section,
Next we replace Assumption~\ref{Jump} by more general assumptions on
the observation kernels and discuss the consequences.

%We first weaken Assumption~\ref{Jump} to the following assumption of

To the best of our knowledge, Assumption~\ref{Vbounded} below, given
first by \cite{Chernoff59}, is the weakest condition in the literature
of hypothesis testing and sequential analysis, and is often interpreted
to an assumption which limits the \textit{excess over the boundary} at
the stopping time \cite{Lorden70}.
%
%There exists $\xi'_M < \infty$ such that

{\renewcommand{\theass}{2$'$}
\begin{ass}\label{Vbounded}
There exists $\xi_M < \infty$ such that
\[
\max_{i,j\in\Omega_M} \max_{a \in\mathcal{A}_M} \int_{\mathcal{Z}}
q_i^a(z) \biggl|\log\frac
{q_i^a(z)}{q_j^a(z)} \biggr|^2 \,dz \le
\xi_M.
\]
\end{ass}}% }}}

%The proofs of Propositions~\ref{propLB1} and~\ref{propUB1} %, and
%Corollary~\ref{corAsym1}
%rely on Chernoff's approach \cite{Chernoff59}, and
%the asymptotic behavior of the bounds remains intact
%if Assumption~\ref{Jump} is replaced with Assumption~\ref{Vbounded}.
%However, as shown in the proof of these propositions in the
%supplemental article~\cite{Naghshvar-SuppA}, Section~\ref{supp:pi1},,
%Assumption~\ref{Jump} allows us to give a precise non-asymptotic
%characterization of the bounds
%by applying the \emph{method of bounded differences}
%and in particular the McDiarmid's inequality~\cite{McDiarmid89}.

Proposition~\ref{propLB1} remains valid even if Assumption~\ref{Jump}
is replaced with Assumption~\ref{Vbounded} (with the only change that
$\xi_M^2$ is replaced with $\xi_M$ in the bound). The proof of
Proposition~\ref{propUB1} relies on Chernoff's approach
\cite{Chernoff59}, and the asymptotic behavior of the bound remains
intact if Assumption~\ref{Jump} is replaced with
Assumption~\ref{Vbounded}. However, as shown in the proof of this
proposition in the supplemental article~\cite{Naghshvar-SuppA},
Section~5, Assumption~\ref{Jump} allows us to give a
precise nonasymptotic characterization of the bound by applying the
\textit{method of bounded differences} and, in particular, McDiarmid's
inequality~\cite{McDiarmid89}.

Next we consider the consequence of weakening Assumption~\ref{Jump} on
Theorems \ref{thmLB2}~and~\ref{thmUB2}, hence on the performance of
policy $\tilde{\pi}_2$. To do so, we consider an even weaker assumption
than Assumption~\ref{Vbounded} as given below:
%Consider the following assumption:
%There exists $\xi''_M < \infty$ such that
%$$\max_{i,j\in\Omega_M} \max_{a \in\mathcal{A}_M}
%D(q^a_i\|q^a_j) \le\xi''_M.$$ % < \infty$.
%%There exists $\xi' < \infty$ such that
%%$$\sup_M \max_{i,j\in\Omega_M} \max_{a \in
%%D(q^a_i\|q^a_j) \le\xi'.$$ % < \infty$.

{\renewcommand{\theass}{2$''$}
\begin{ass}\label{KLbounded}
There exist $\xi_M < \infty$ and $\gamma>0$ such that
\[
\max_{i,j\in\Omega_M} \max_{a \in\mathcal{A}_M} \int_{\mathcal{Z}}
q_i^a(z) \biggl|\log\frac
{q_i^a(z)}{q_j^a(z)} \biggr|^{1+\gamma} \,dz \le
\xi_M.
\]
%
%{\textit{There exists $\xi_M < \infty$ such that
%$$D_{\max}(M) \le\xi_M.$$ % < \infty$.
%$$\max_{i,j\in\Omega_M} \max_{a \in\mathcal{A}_M}
%D(q^a_i\|q^a_j) \le\xi_M.$$ % < \infty$.
\end{ass}}

Define function $\psi_M\dvtx \mathbb{R}_+\to\mathbb{R}_+$ as follows:
\[
%_{a \in\mathcal{A}_M}
\psi_M(b):= \max_{i,j\in\Omega_M} \max_{a \in
\mathcal{A}_M}\int
_{\mathcal{Z}} q_i^a(z) \biggl[\log
\frac
{q_i^a(z)}{q_j^a(z)} \biggr]_b \,dz,
\]
where $[g]_b=g {\mathbf{1}}_{\{g>b\}}$. Note that $\psi_M(b)$ is in
general nonincreasing in $b$, and if Assumption~\ref{KLbounded} holds,
$\psi_M(b) \le b^{-\gamma} \xi_M$.
%We have the following lower and upper bounds:
Under the weaker Assumption~\ref{KLbounded} (and naturally
Assumption~\ref{Vbounded}), Theorems~\ref{thmLB2} and~\ref{thmUB2} can
be replaced by the following:

%
%pr3 #&#
\begin{proposition}
\label{propLB2gen} Under Assumptions~\ref{KL0} and~\ref{KLbounded} and
for ${L}> \frac {\log M}{I_{\max}(M)}$, $\bolds{\rho} \in\mathbb
{P}_{{L}}(\Omega_M)$, $\delta\in(0,0.5]$, and $b>0$,
\begin{eqnarray*}
V^*(\bolds{\rho}) \ge \underline{V_{3}}(\bolds{\rho})
&:= & \frac{1}{1+\psi_M(b)/D_{\max}(M)}
\\
&&{}\times  \biggl[ \frac
{H(\bolds{\rho}) - H([\delta,1-\delta]) - {\delta} \log
(M-1)}{I_{\max}(M)}
\\
&&\hspace*{17pt}{} + \frac{\log ((1-{L}^{-1})/L^{-1}) - \log
((1-\delta)/\delta)-b}{D_{\max}(M)}
\\
&&\hspace*{105pt}
{}\times \mathbf{1}_{\{\max_{i\in\Omega_M} \rho_i \le1 - \delta\}} - K'_3 \biggr]^+,
\end{eqnarray*}
%
%is a lower bound for the optimal value function $V^{*}$
where $K'_3$ is a positive constant independent of $\delta$ and ${L}$.
In addition, if Assumption~\ref{JumpM} also holds, then $K'_3$ can be selected
independent of $M$ as well.
\end{proposition}

The proof is provided in the supplemental
article~\cite{Naghshvar-SuppA}, Section~8.1. %\ref{supp:LB2gen}.

%%%\ignore{
%%%%
%%%\begin{remarks}
%%%If Assumption~\ref{Jump} holds, then we have $\psi_M(b)=0$ for $b\ge
%%%\xi_M$. Therefore, under Assumption~\ref{Jump} (and setting $b=\xi_M$),
%%%lower bound $\underline{V_{3}}$ simplifies to $\underline{V_2}$ given
%%%in Theorem~\ref{thmLB2}.
%%%\end{remarks}
%%%%
%%%}

%We can also obtain the following upper bound:

%
%pr4 #&#
\begin{proposition}
\label{propUB2gen} Under Assumptions~\ref{KL0} and~\ref{KLbounded}, and
for ${L}> 1$ and $\bolds{\rho} \in\mathbb{P}_{{L}}(\Omega_M)$, $\exists
b'\in (0,\infty)$ such that for all $b\ge b'$, $0
\le\frac{(1+(\log e)/b)2^{-b}\psi_M(b)}{I_2(M)-\psi_M(b)}<1$,
and\looseness=-1
\begin{eqnarray*}
V^*(\bolds{\rho}) &\le&\widebar{V}_{3} (\bolds{\rho})
\\
&:=& %\frac{1}{1-\frac{(1+\frac{1}{b})2^{-b}\psi_M(b)}{I_2(M)-\psi_M(b)}} \times\\
\biggl(1-\frac{(1+(\log e)/b)2^{-b}\psi_M(b)}{I_2(M)-\psi
_M(b)} \biggr)^{-1}
\\
&&{}\times
\Biggl(\frac{H(\bolds{\rho})+\log (\tilde{\rho
}/(1-\tilde{\rho}))+b+\log e}{I_2(M)-\psi_M(b)} + %\sum_{i=1}^M \rho_i \frac{\log\frac{1-\L^{-1}}{\L^{-1}}}{D_{
\sum
_{i=1}^M \rho_i
\frac{\log{L}}{D_{\bolds{\eta}_i}(M)-\psi_M(b)} \Biggr)
\\
&&{}+1.
\end{eqnarray*}\looseness=0
%
%is an upper bound for the optimal value function $V^{*}$.
\end{proposition}

The proof is provided in the supplemental
article~\cite{Naghshvar-SuppA}, Section~8.2. %\ref{supp:UB2gen}.

%Note that if Assumption~\ref{Jump} holds, then we have $\psi_M(b)=0$
%for $b\ge\xi_M$.
%%Therefore, under Assumption~\ref{Jump} (and setting $b=\xi_M$),
%%$\underline{V_{3}}$ and $\widebar{V}_{3}$ simplify to
%%$\underline{V_2}$ and $\widebar{V}_{2}$, respectively.
%
%In general, $\psi_M(b)$ is non-increasing in $b$;
%and under Assumption~\ref{KLbounded}, $\psi_M(0) \le\xi_M+\frac{\log
%e}{e}$,
%hence, $\psi_M(b)\to0$ as $b\to\infty$.
%Under Assumption~\ref{KLbounded}, $\psi_M(0) \le\xi_M+\frac{\log
%e}{e}$,
%and hence $\psi_M(b)\to0$ as $b\to\infty$ %(since the tail of a
%convergent series tends to 0).
%(since $\psi_M(0)$ can be written as a convergent series with the tail
%$\psi_M(b)$).
%Furthermore, if Assumption~\ref{JumpM} holds,
%then %$\sup_M \psi_M(0) \le\xi''+\frac{\log e}{e}$, and
%$\sup_M \psi_M(b)\to0$ as $b\to\infty$.
As we discussed, $\psi_M(b) \le b^{-\gamma}\xi_M$ under
Assumption~\ref{KLbounded}. Furthermore, if Assumption~\ref{JumpM}
holds, then $\sup_M \psi_M(b) \le b^{-\gamma}\xi$. In other words, we
can select $b$ as a function of ${L}$ and $M$ (e.g., $b=\log\log{L}M$)
such that $\underline{V_{3}}$ and $\widebar{V}_{3}$ have the same
dominating terms (in ${L}$ and $M$) as $\underline{V_2}$ and
$\widebar{V}_{2}$, respectively.\vadjust{\goodbreak}

In summary, the asymptotic results of the paper presented in
Section~\ref{Optimality}
%hold even under the weaker Assumption~2''
hold under the weaker Assumptions \ref{Vbounded} and
\ref{KLbounded} replacing Assumption~\ref{Jump} (with the only exception
that the asymptotic optimality of order-2 of policy $\tilde{\pi}_2$
established in Corollary~\ref{corAsym1} is degraded to asymptotic
optimality of order-1). Our choice to present the work under
Assumption~\ref{Jump}, however, significantly simplifies the
presentation and also enables a precise nonasymptotic characterization
%%of the non-dominant terms in
of the lower and upper bounds.

%%%%--------------------------------------------------------------------------------
%s8 #&#
\section{Conclusions and future work}\label{Discussion}

In this paper we considered the problem of active sequential $M$-ary
hypothesis testing.
Using a DP formulation, we characterized the optimal value function $V^*$.
Three lower bounds (complementary for various values of the parameters
of the problem) were obtained for the optimal value function $V^*$.
We also proposed two heuristic policies whose performance analysis
resulted in two upper bounds for $V^*$.
Subsequently, we discussed important consequences of the bounds and
established order and asymptotic optimality of the proposed policies
under different scenarios.
An important problem which remains is further improvement of the
performance bounds.

In this paper we focused on sequential policies, that is, policies
whose sample size is not known initially and is dependent on the
observation outcomes. There exist other types of policies in the
literature. For example, nonsequential policies take a fixed number of
samples (independent of observation outcomes) and make the final
decision afterward, while multi-stage policies (introduced
in~\cite{Lorden83,Bartroff07}) can take a retire--declare action only at
the end of each stage, and stages are not necessarily of the same size.
Comparing the performance of sequential, nonsequential and multi-stage
policies in the context of active hypothesis testing is an area of
future work.

In this paper we assumed that all sensing actions incur one unit of
cost (each action can be executed in one unit of time). It is also of
interest to consider the scenario where there is a cost associated with
each action which, for example, characterizes the amount of energy or
time required to perform that action; and the goal is to find the true
hypothesis subject to a cost criterion. Such generalization has been
studied for the problem of variable-length coding with feedback
in~\cite{Nakiboglu08}.

%{\allowdisplaybreaks[4]{
%%%%Appendix--------------------------------------------------------------------------------------
\begin{appendix}\label{AppLB}
\section*{Appendix: Proof of Theorems 1--3}
%, \ref{thmLB2}, and \ref{thmUB2}} % and Corollary~\ref{PeTo1}}

%s9.1 #&#
\subsection{\texorpdfstring{Proof of Theorem~\protect\ref{thmLB}}{Proof of Theorem 1}}
\label{AppLB1}

Let ${\Gamma}$ be the set of all mappings $\gamma\dvtx  \Omega_M \to
\Omega_M$ such that $\gamma(i) \neq i$ for $i \in\Omega_M$. Now
associated with any $\gamma\in{\Gamma}$, define
%e33 #&#
\begin{equation}
\underline{V_1}^{\gamma}(\bolds{\rho})= \Biggl[ \sum
_{i=1}^{M} \rho_i
\frac{\log((1-L^{-1})/L^{-1}) - \log( \rho_i/\rho_{\gamma
(i)})}{\max_{\hat{a}\in\mathcal{A}_M} D(q^{\hat{a}}_i\|
q^{\hat{a}}_{\gamma(i)})} - K'_1 \Biggr]^{+}.\vadjust{\goodbreak}
\end{equation}
Next we use Lemma~\ref{lemVLB} to show that $V^* \ge\underline
{V_1}^{\gamma}$ for all $\gamma\in\Gamma$. In particular, we show that
for all $\gamma\in\Gamma$ and all $\bolds{\rho}
\in\mathbb{P}(\Omega_M)$, $\underline{V_1}^{\gamma}(\bolds{\rho})
\le\min\{ 1+\min_{a \in\mathcal{A}_M}(\mathbb{T}^{a} \underline
{V_1}^{\gamma})(\bolds{\rho}), \min_{j \in\Omega_M} (1-\rho_j) {L}\}$.
For any $\bolds{\rho}$ such that $\underline{V_1}^{\gamma
}(\bolds{\rho})=0$, the inequality holds trivially. For
$\underline{V_1}^{\gamma}(\bolds{\rho})>0$ and for any action $a
\in\mathcal{A}_M$, we have
\begin{eqnarray*}
&& \bigl(\mathbb{T}^{a} \underline{V_1}^{\gamma}
\bigr) (\bolds{\rho})
\\
&&\qquad \ge\sum_{i=1}^{M}
\int\rho_i q^a_i(z) \frac{\log((1-L^{-1})/L^{-1}) - \log(\rho_i q^a_i(z)/(\rho
_{\gamma(i)} q^a_{\gamma(i)}(z)))} {
\max_{\hat{a}\in\mathcal{A}_M} D(q^{\hat{a}}_i\|q^{\hat
{a}}_{\gamma(i)})} \,dz
\\
&&\quad\qquad{}  - K'_1
\\
&&\qquad = \underline{V_1}^{\gamma}(\bolds{\rho})- \sum
_{i=1}^{M} \rho_i \frac{D(q^a_i\|q^a_{\gamma(i)})}{\max_{\hat{a}\in\mathcal{A}_M} D(q^{\hat{a}}_i\|q^{\hat{a}}_{\gamma
(i)})}
\\
&&\qquad \ge\underline{V_1}^{\gamma}(\bolds{\rho}) - 1.
\end{eqnarray*}

%
%cl1 #&#
\begin{claim}[(In Section~9.1 of the supplemental
article~\cite{Naghshvar-SuppA})] %[in Appendix~\ref{Constants}]
\label{LBK1} Constant $K'_1$ can be selected independent of ${L}$ such
that $\underline{V_1}^{\gamma}(\bolds{\rho}) \le\min_{j \in \Omega_M}
(1-\rho_j) {L}$ is satisfied for all $\gamma\in\Gamma$.
\end{claim}
%
%The proof of Claim~\ref{LBK1} is provided in Appendix~\ref{Constants}.

Using Claim~\ref{LBK1} and letting $\underline{V_1}(\cdot) =
\max_{\gamma\in\Gamma} \underline{V_1}^{\gamma}(\cdot)$, we have the
assertion of the theorem.

%%%%%%%%%%%%%%%%%%%%%%%%%%%%%%%%%%%%%%%%%%%%%%%%%%%%%%%%
%%%%%%%%%%%%%%%%%%%%%%%%%%%%%%%%%%%%%%%%%%%%%%%%%%%%%%%%

%s9.2 #&#
\subsection{\texorpdfstring{Proof of Theorem~\protect\ref{thmLB2}}{Proof of Theorem 2}}
\label{AppLB2}

%Recall that $I_{\max} = \max_{a\in\mathcal{A}_M} \max
%_{\hat{\bolds{\rho}}\in\mathbb{P}(\Omega_M)}
%I(\hat{\bolds{\rho}};q^{a}_{\hat{\bolds{\rho}}})$
%and $\alpha(\L,M)=\frac{M-1}{M-1+2^{\L I_{\max}}}$.
%
We first show that for all $\bolds{\rho} \in\mathbb{P}(\Omega
_M)$, %belief vectors $\bolds{\rho} \in\mathbb{P}(\Omega_M)$,
%
%e34 #&#
\begin{eqnarray}
%V^*(\bolds{\rho}) &\ge
%- \alpha(\L,M) \log(M-1)}{I_{\max}} +
V^*(
\bolds{\rho}) &\ge& \biggl[ \frac{H(\bolds{\rho}) - H([\alpha({L},M),1-\alpha
({L},M)]) - \alpha({L},M) \log(M-1)}{I_{\max}(M)}\hspace*{-30pt}
\nonumber\\[-15pt]\label{LBG}  \\[-5pt]
&&\hspace*{218pt}{}+ \alpha({L},M) {L} \biggr]^+\!.\nonumber\hspace*{-28pt}
\end{eqnarray}
Note that the right-hand side of~(\ref{LBG}) can be written as
%
%e35 #&#
\begin{equation}
G(\bolds{\rho}):= \biggl[\frac{H(\bolds{\rho}) -
H(\bolds{\nu})}{I_{\max}(M)} + \alpha({L},M) {L} \biggr]^+,
\end{equation}
where
%e36 #&#
\begin{equation}
\bolds{\nu}= \biggl[\frac{\alpha({L},M)}{M-1},\ldots,\frac
{\alpha({L},M)}{M-1},1-
\alpha({L},M) \biggr].
\end{equation}
Next we show that $G(\bolds{\rho}) \le\min\{ 1+\min_{a
\in\mathcal{A}_M}(\mathbb{T}^{a} G)(\bolds{\rho}), \min_{j \in\Omega_M}
(1-\rho_j) {L}\}$ for all $\bolds{\rho} \in\mathbb{P}(\Omega_M)$. For
any $\bolds{\rho}$ such that $G(\bolds{\rho})=0$, the inequality holds
trivially. For $G(\bolds{\rho})>0$ and for any action $a \in\mathcal
{A}_M$, we have
%
%e37 #&#
\begin{eqnarray}
\bigl(\mathbb{T}^{a}G\bigr) (\bolds{\rho}) &=&
\frac{\int H(\bolds{\Phi}^a(\bolds{\rho},z)) q^a_{\bolds{\rho}}(z) \,dz -
H(\bolds{\nu})} {I_{\max}(M)} + \alpha({L},M) {L}\nonumber
\\
&=& \frac{H(\bolds{\rho}) - I(\bolds{\rho};q^{a}_{\bolds{\rho}}) - H(\bolds{\nu})} {
I_{\max}(M)} + \alpha({L},M) {L}
\nonumber\\[-8pt]\\[-8pt]
&=& G(\bolds{\rho}) - \frac{I(\bolds{\rho};q^{a}_{\bolds{\rho}})}{I_{\max}(M)}\nonumber
\\
&\ge& G(\bolds{\rho}) - 1,\nonumber
\end{eqnarray}
where the last inequality follows from the fact that
\[
I\bigl(\bolds{\rho};q^{a}_{\bolds{\rho}}\bigr) \le\max_{\hat{a}\in\mathcal{A}_M}
\max_{\hat{\bolds{\rho}}\in\mathbb{P}(\Omega_M)} I\bigl(\hat{\bolds{\rho}};q^{\hat
{a}}_{\hat{\bolds{\rho}}}
\bigr) = I_{\max}(M).
\]
Therefore,
\[
G(\bolds{\rho}) \le1+\min_{a \in\mathcal
{A}_M}\bigl(\mathbb{T}^{a} G\bigr)
(\bolds{\rho}).
\]

What remains is to show that $G(\bolds{\rho}) \le\min_{j\in\Omega_M}
(1-\rho_j) {L}$. Rewriting $G$ as
\begin{eqnarray*}
G(\bolds{\rho}) &=& \biggl[\frac{\sum_{i=1}^{M-1} \rho_i \log (1/\rho_i) +
(1-\sum_{i=1}^{M-1} \rho_i) \log (1/(1-\sum_{i=1}^{M-1} \rho
_i)) - H(\bolds{\nu})}{I_{\max}(M)}
\\[-4pt]
&&\hspace*{251pt}{}
 + \alpha({L},M) {L} \biggr]^+\!,
\end{eqnarray*}
we can compute the gradient at $\bolds{\nu}$. For all
$i=1,2,\ldots,M-1$,
\begin{eqnarray*}
\frac{\partial G}{\partial\rho_i}(\bolds{\nu}) &=& \biggl( \log\frac{1}{\rho_i} - \log e -
\log\frac{1}{1-\sum_{i=1}^{M-1} \rho_i} + \log e \biggr) \Big/ I_{\max}(M) \bigg|_{\bolds{\rho}=\bolds{\nu}}
\\
&=& \biggl( \log\frac{\rho_M}{\rho_i} \biggr) \Big/ I_{\max}(M)
\Big|_{\bolds{\rho}=\bolds{\nu}} %\\
= \biggl( \log\frac{1-\alpha({L},M)}{\alpha({L},M)/(M-1)} \biggr) \Big/
I_{\max}(M) %\\
%&= \left( \log\frac{1-\frac{M-1}{M-1+2^{\L I_{\max}}}}{
= {L}.
\end{eqnarray*}
Furthermore, $G(\bolds{\nu})= \alpha({L},M) {L}= (1-\nu_M) {L}$.
Without loss of generality and since both functions $G(\bolds{\rho})$
and $\min_{j\in\Omega_M} (1-\rho_j) {L}$ are symmetric, let us focus on
$\mathbb{P}_M(\Omega_M):=  \{ \bolds{\rho} \in\mathbb
{P}(\Omega_M)\dvtx  \rho_M \ge\rho_i,\ \forall i \in\Omega_M - \{M\}
\}$. In this case, $\min_{j \in\Omega_M} (1-\rho_j) {L}= (1-\rho _M)
{L}= \sum_{i=1}^{M-1}\rho_i {L}$ and, hence, $\min_{j \in\Omega_M}
(1-\rho_j) {L}$ is the tangent hyperplane to $G(\bolds{\rho})$ at
$\bolds{\nu}$. This along with concavity of function $G$ implies
$G(\bolds{\rho}) \le\min_{j \in\Omega_M} (1-\rho_j) {L}$. Using
Lemma~\ref{lemVLB}, we have the assertion of the theorem.

%%%%%%%%%%%%%%%%%%%%
%%%%%%%%%%%%%%%%%%%%
Next we need to show that
%e38 #&#
\begin{eqnarray}
\label{V2def}
V^*(\bolds{\rho}) \ge \underline{V_2}(
\bolds{\rho})
&= & \biggl[ \frac{H(\bolds{\rho}) - H([\delta,1-\delta]) -
{\delta} \log(M-1)}{I_{\max}(M)}\nonumber
\\
&&\hspace*{4pt}
{} + \frac{\log ((1-L^{-1})/L^{-1}) -
\log ((1-\delta)/\delta)-\xi_M}{D_{\max}(M)}
\\[-4pt]
&&\hspace*{99pt}
{}\times {\mathbf{1}}_{\{\max_{i\in\Omega_M} \rho_i \le1 -
\delta\}} - K'_2 \biggr]^+.\nonumber
\end{eqnarray}
We show this in two steps. First we consider the following function:
%e39 #&#
\begin{eqnarray}
\label{JDef} J'(\bolds{\rho})&:=& \Biggl[ \sum
_{i=1}^{M} \rho_i \frac{\log((1-L^{-1})/L^{-1}) - \log (\rho_i/(1-\rho
_i))}{D_{\max}(M)} -
K'_2 \Biggr]^+.
\end{eqnarray}
%
%where $\xi:= \max_{i,j\in\Omega_M} \max_{a \in
%and $\xi< \infty$ by Assumption~\ref{Jump}.
We use Jensen's inequality to show that
%for all $\bolds{\rho}\in\mathbb{P}(\Omega_M)$,
%e40 #&#
\begin{equation}
\label{ineqJ} J'(\bolds{\rho}) \le1+\min_{a \in\mathcal
{A}_M}\bigl(
\mathbb{T}^{a} J'\bigr) (\bolds{\rho})\qquad\forall
\bolds{\rho}\in\mathbb{P}(\Omega_M).
\end{equation}
%
%For all $\bolds{\rho}\in\mathbb{P}(\Omega_M)$,
%$J'(\bolds{\rho}) \le1+\min_{a \in\mathcal{A}_M}(
%
For any $\bolds{\rho}$ such that $J'(\bolds{\rho})=0$, inequality
(\ref{ineqJ}) holds trivially. For any $\bolds{\rho}$ such that
$J'(\bolds{\rho})>0$ and for any $a \in\mathcal{A}_M$, we have
\begin{eqnarray*}
&& \bigl(\mathbb{T}^{a}J'\bigr) (\bolds{\rho})
\\
&&\qquad \ge \sum_{i=1}^{M} \int\rho_i
q^a_i(z) \frac{\log ((1-L^{-1})/L^{-1}) - \log(\rho_i q^a_i(z)/\sum_{j
\neq i} \rho_j q^a_j(z))}{D_{\max}(M)} \,dz
\\[-1pt]
&&\quad\qquad{} - K'_2
\\[-3pt]
&&\qquad = J'(\bolds{\rho})- \sum_{i=1}^{M}
\rho_i \frac{\int q^a_i(z) \log(q^a_i(z)/\sum_{j \neq i} (\rho_j/(1-\rho_i)) q^a_j(z)) \,dz}{D_{\max}(M)}
\\
&&\qquad \ge J'(\bolds{\rho})- \sum_{i=1}^{M}
\rho_i \frac{\sum_{j\neq i} (\rho_j/(1-\rho
_i)) D(q^a_i\|q^a_j)}{D_{\max}(M)}
\\
&&\qquad \ge J'(\bolds{\rho}) - 1.
\end{eqnarray*}

Next we define $J(\bolds{\rho})=\max\{J'(\bolds{\rho}),
J''(\bolds{\rho})\}$, where $J''(\bolds{\rho})$ is the right-hand side
of~(\ref{V2def}), that is,
%
%e41 #&#
\begin{eqnarray}
\label{JDef} J''(\bolds{\rho}) &= & \biggl[
\frac{H(\bolds{\rho}) - H([\delta,1-\delta]) -
{\delta} \log(M-1)}{I_{\max}(M)}\nonumber
\\
&&\hspace*{4pt}{} + \frac{\log((1-L^{-1})/L^{-1}) -
\log((1-\delta)/\delta)-\xi_M}{D_{\max}(M)}
\\[-4pt]
&&\hspace*{100pt}
{}\times \mathbf{1}_{\{\max_{i\in\Omega_M} \rho_i \le1 - \delta\}} -K'_2 \biggr]^+.\nonumber
\end{eqnarray}
\begin{itemize}%[\textit{Case} 2:]
\item \textit{Case} 1: For all $\bolds{\rho}$ such that
    $J(\bolds{\rho})=0$ or $J(\bolds{\rho})=J'(\bolds{\rho})$, it
    is trivial from (\ref{ineqJ}) that
%e42 #&#
\begin{equation}
\label{VLB2case1} J(\bolds{\rho}) = J'(\bolds{\rho}) \le1+
\min_{a
\in\mathcal{A}_M}\bigl(\mathbb{T}^{a} J'\bigr) (
\bolds{\rho}) \le1+\min_{a \in\mathcal{A}_M}\bigl(\mathbb{T}^{a} J\bigr) (
\bolds{\rho}).
\end{equation}

\item \textit{Case} 2: For all $\bolds{\rho}$ such that
    $J(\bolds{\rho})=J''(\bolds{\rho})>0$, and for any action $a
    \in\mathcal{A}_M$, we have
%
% {\allowdisplaybreaks{
%e43 #&#
\begin{eqnarray}\label{VLB2case2}
\bigl(\mathbb{T}^{a}J\bigr) (\bolds{\rho})\nonumber
&=& \int J\bigl(\bolds{\Phi}^a(\bolds{\rho},z)\bigr)
q^a_{\bolds{\rho}}(z) \,dz\nonumber
\\
&\stackrel{(a)} {\ge}& \frac{\int H(\bolds{\Phi}^a(\bolds{\rho},z)) q^a_{\bolds{\rho}}(z) \,dz - H([\delta,1-\delta]) -
{\delta} \log(M-1)}{I_{\max}(M)}\nonumber
\\
&&{} + \frac{\log((1-L^{-1})/L^{-1}) - \log
((1-\delta)/\delta)-\xi_M}{D_{\max}(M)}\nonumber
\\
&&\quad{} \times \mathbf{1}_{\{\max_{i\in\Omega_M} \rho_i \le1 -\delta\}}- K'_2
\\
&=& J''(\bolds{\rho}) -
\frac{I(\bolds{\rho};q^{a}_{\bolds{\rho}})}{I_{\max}(M)}\nonumber
\\
&\ge& J''(\bolds{\rho}) - 1\nonumber
\\
&\stackrel{(b)} {=}& J(\bolds{\rho}) - 1,\nonumber
\end{eqnarray}
%
%}}
where ($a$) follows from Claim~\ref{claimJ} below and ($b$) holds
since $\bolds{\rho}$ is such that
$J(\bolds{\rho})=J''(\bolds{\rho})$.
\end{itemize}

%
%cl2 #&#
\begin{claim}[(In Section~9.2 of the supplemental
article~\cite{Naghshvar-SuppA})] %[in Appendix~\ref{Constants}]
\label{claimJ} Let $\bolds{\rho}$ be such that
$J(\bolds{\rho})=J''(\bolds{\rho})>0$. If Assumption~\ref{Jump} holds,
then for all actions $a\in\mathcal {A}_M$ and observations
$z\in\mathcal{Z}$,
%e44 #&#
\begin{eqnarray}\label{IneqJ}
J\bigl(\bolds{\Phi}^a(\bolds{\rho},z)\bigr)
&\ge& \frac{H(\bolds{\Phi}^a(\bolds{\rho},z)) -
H([\delta,1-\delta]) - {\delta} \log(M-1)}{I_{\max}(M)}\nonumber
\\
&&{}+ \frac{\log((1-L^{-1})/L^{-1}) - \log
((1-\delta)/\delta)-\xi_M}{D_{\max}(M)}
\\
&&\quad{}\times \mathbf{1}_{\{\max_{i\in\Omega_M} \rho_i \le1 -
\delta\}}- K'_2.\nonumber
\end{eqnarray}
\end{claim}

%In other words, from~\eqref{VLB2case1} and~\eqref{VLB2case2}, we have
%that
Combining~(\ref{VLB2case1}) and~(\ref{VLB2case2}), we have that
%e45 #&#
\begin{equation}
\label{JineqT} J(\bolds{\rho}) \le1+\min_{a \in\mathcal
{A}_M}\bigl(\mathbb{T}^{a}
J\bigr) (\bolds{\rho}).
\end{equation}
We also have
the following:%

%cl3 #&#
\begin{claim}[(In Section~9.3 of the supplemental
article~\cite{Naghshvar-SuppA})] %[in Appendix~\ref{Constants}]
\label{LBK2} For ${L}>\frac{\log M}{I_{\max}(M)}$, constant $K'_2$ can
be selected independent of $\delta$ and ${L}$ such that
$J(\bolds{\rho}) \le\min_{j \in\Omega_M} (1-\rho_j) {L}$. Furthermore,
if $\sup_M \xi_M < \infty$, then $K'_2$ can be selected independent of
$M$ as well.
%For $\L> \frac{\log M}{I_{\max}(M)}$, constant $K'_2$ can be
%selected independent of $\L$ and $M$ such that
%$J(\bolds{\rho}) \le\min_{j \in\Omega_M} (1-\rho_j)
\end{claim}
%
%The proof of Claim~\ref{LBK2} is provided in Appendix~\ref{Constants}.

Lemma~\ref{lemVLB}, together with~(\ref{JineqT}) and Claim~\ref{LBK2},
implies that $V^* \ge J =\break \max\{J', J''\} \ge J'' = \underline{V_2}$.
This is a slightly stronger result than~(\ref{V2def}).

\subsection{\texorpdfstring{Proof of Theorem~\protect\ref{thmUB2}}{Proof of Theorem 3}}
\label{AppUB2}

Recall that $\rho_i(n)$ denotes the posterior belief about
hypothesis~$H_i$ after $n$ observations.\vadjust{\goodbreak}
Let $\tau$, $\tau_i$, $i\in\Omega_M$, be Markov stopping times
defined as follows:
%e46 #&#
%e47 #&#
\begin{eqnarray}
\label{TauDef} \tau&:=&\min \Bigl\{ n\dvtx  \max_{j\in\Omega_M}
\rho_j(n) \ge 1-{L}^{-1} \Bigr\},
\\
\label{TauiDef} \tau_i &:=&\min \bigl\{ n\dvtx  \rho_i(n)
\ge1-{L}^{-1} \bigr\}.
\end{eqnarray}
From~(\ref{Objective2}), the expected total cost under policy $\tilde
{\pi}_2$ is upper bounded as
%e48 #&#
\begin{eqnarray}
V_{\tilde{\pi}_2}(\bolds{\rho}) &=&
\mathbb{E}_{\tilde{\pi
}_2}\Bigl[\tau+ \min_{j\in\Omega_M} \bigl(1-
\rho_j(\tau)\bigr) {L}\Bigr]\nonumber
\\
&\le&\mathbb{E}_{\tilde{\pi}_2}[\tau] + 1\label{UBVpi2}
\\
& \le&\sum_{i=1}^{M}
\rho_i \mathbb{E}_{\tilde{\pi}_2}[\tau_i | \theta=i] +
1,\nonumber
\end{eqnarray}
where $\bolds{\rho}=[\rho_1,\rho_2, \ldots, \rho_M]=[\rho
_1(0),\rho_2(0), \ldots,\rho_M(0)]$ %is the prior belief about
%$H_1,H_2,\ldots,H_M$
and the last inequality follows from the fact that $\tau\leq\tau_i$,
$\forall i\in\Omega_M$.
For notational simplicity, subscript $\tilde{\pi}_2$ is dropped for
the rest of the proof.

Next we find an upper bound for $\mathbb{E}[\tau_i | \theta=i]$,
$i\in\Omega_M$.
Let
%e49 #&#
\begin{equation}
\label{UDef} U_n:= \log\frac{\rho_i(n)}{1-\rho_i(n)} - \log
\frac{\tilde{\rho
}}{1-\tilde{\rho}}
\end{equation}
and let $\mathcal{F}_n$ denote the history of previous actions and
observations up to time $n$, that is, $\mathcal{F}_n:= \sigma\{
\bolds{\rho}(0), A(0), Z(0), \ldots, A(n-1), Z(n-1) \}$. Under policy
$\tilde{\pi}_2$, the sequence $\{U_n\}$, $n=0,1,\ldots,$ forms a
submartingale with respect to the filtration $\{\mathcal{F}_n\} $ with
the following properties:
\begin{longlist}[(C3)]
\item[(C1)] If $U_n < 0$ and $\rho_j(n) < \tilde{\rho}$ for all
    $j\in\Omega_M$ ($\Rightarrow P(A(n)=a)=\eta_{0a}$):
\begin{eqnarray*}
&& \mathbb{E} [U_{n+1} - U_n |
\mathcal{F}_n, \theta= i ]
\\
&& \qquad = \sum_{a \in\mathcal{A}_M} P\bigl(A(n)=a\bigr) \mathbb{E}
\bigl[U_{n+1} - U_n | \mathcal{F}_n, \theta=
i, A(n)=a \bigr]
\\
&&\qquad = \sum_{a \in\mathcal{A}_M} \eta_{0a} \mathbb{E}
\bigl[U_{n+1} - U_n | \mathcal{F}_n, \theta=
i, A(n)=a \bigr]
\\
&&\qquad = \sum_{a \in\mathcal{A}_M} \eta_{0a} \mathbb{E} \biggl[
\log \frac{\rho_i(n) q^{a}_i(Z)}{\sum_{j \neq i} \rho_j(n) q^{a}_j(Z)} - \log\frac{\rho_i(n)}{1-\rho_i(n)} \bigg| \mathcal{F}_n,
\theta= i \biggr]
\\
&&\qquad = \sum_{a \in\mathcal{A}_M} \eta_{0a} \int
q^a_i(z) \log\frac
{q^a_i(z)}{\sum_{j \neq i} (\rho_j(n)/(1-\rho_i(n)))
q^a_j(z)}\,dz
\\
&&\qquad \ge\max_{\bolds{\lambda} \in\mathbb{P}(\mathcal
{A}_M)}\ \min_{i\in\Omega_M}\ \min_{\hat{\bolds{\rho}}\in\mathbb{P}_{{L}}(\Omega_M)} \sum_{a \in\mathcal
{A}_M}
\lambda_{a} D\biggl(q^{a}_i\bigg\| \sum
_{j \neq i} \frac{{\hat
{\rho}}_j}{1-\hat{\rho}_i} q^{a}_j\biggr)
\\
&&\qquad = I_{\bolds{\eta}_0}(M).
\end{eqnarray*}

If $U_n < 0$ and $\rho_k(n) \ge\tilde{\rho}$ for some $k \neq i$
($\Rightarrow P(A(n)=a)=\eta_{ka}$):
\begin{eqnarray*}
&& \mathbb{E} [U_{n+1} - U_n |
\mathcal{F}_n, \theta= i ]
\\
%&= \sum_{a \in\mathcal{A}_M} P(A(n)=a) \mathbb{E} \left[U_{n+1} -
%U_n | \mathcal{F}_n, \theta= i, A(n)=a \right]\\
&&\qquad = \sum_{a \in\mathcal{A}_M}
\eta_{ka} \mathbb{E} \bigl[U_{n+1} - U_n |
\mathcal{F}_n, \theta= i, A(n)=a \bigr]
\\
%&= \sum_{a \in\mathcal{A}_M} \eta_{0a} \mathbb{E} \left[\log\frac{
&&\qquad = \sum_{a \in\mathcal{A}_M}
\eta_{ka} \int q^a_i(z) \log
\frac
{q^a_i(z)}{\sum_{j \neq i} (\rho_j(n)/(1-\rho_i(n)))
q^a_j(z)}\,dz
\\
&&\qquad \ge\min_{i\in\Omega_M} \min_{k \neq i} \min_{\hat{\bolds{\rho}}\dvtx  \hat{\rho}_k\ge\tilde{\rho}} \sum
_{a \in\mathcal{A}_M}
\eta_{ka} D\biggl(q^{a}_i\bigg\| \sum
_{j \neq i} \frac{{\hat{\rho}}_j}{1-\hat{\rho}_i} q^{a}_j\biggr)
\\
&&\qquad = I_{\bolds{\eta},\tilde{\rho}}(M);
\end{eqnarray*}

\item[(C2)] If $U_n \ge0$ ($\rho_i(n) \geq\tilde{\rho} \Rightarrow
    P(A(n)=a)=\eta_{ia}$):
%
%{\allowdisplaybreaks{
\begin{eqnarray*}
&& \mathbb{E} [U_{n+1} - U_n |
\mathcal{F}_n, \theta= i ]
\\
%&= \sum_{a \in\mathcal{A}_M} P(A(n)=a) \mathbb{E} \left[U_{n+1} -
%U_n | \mathcal{F}_n, \theta= i, A(n)=a \right]\\
&&\qquad = \sum_{a \in\mathcal{A}_M}
\eta_{ia} \mathbb{E} \bigl[U_{n+1} - U_n |
\mathcal{F}_n, \theta= i, A(n)=a \bigr]
\\
%&= \sum_{a \in\mathcal{A}_M} \eta_{ia} \mathbb{E} \left[\log\frac{
&&\qquad = \sum_{a \in\mathcal{A}_M}
\eta_{ia} \int q^a_i(z) \log
\frac{q^a_i(z)}{\sum_{j \neq i} (\rho_j(n)/(1-\rho_i(n)))
q^a_j(z)}\,dz
\\
&&\qquad \ge \max_{\bolds{\lambda} \in\mathbb{P}(\mathcal
{A}_M)} \min_{\hat{\bolds{\rho}}\in\mathbb
{P}_{{L}}(\Omega_M)} \sum_{a \in\mathcal{A}_M}
\lambda_{a} D\biggl(q^{a}_i\bigg\| \sum
_{j \neq i} \frac{{\hat{\rho}}_j}{1-\hat
{\rho}_i} q^{a}_j\biggr)
\\
&&\qquad = D_{\bolds{\eta}_i}(M);
\end{eqnarray*}

\item[(C3)] $|U_n-U_{n-1}| \leq\max_{i,j\in\Omega_M} \max
    _{a\in\mathcal{A}_M} \sup_{z\in\mathcal{Z}} \log
    \frac{q^a_i(z)}{q^a_j(z)} \le\xi_M$.
\end{longlist}

Stopping time $\tau_i$ defined in (\ref{TauiDef}) can be rewritten as
%e50 #&#
\begin{eqnarray}\label{tauredef}
\qquad \tau_i &=&\min \bigl\{ n\dvtx  \rho_i(n) \ge1-{L}^{-1} \bigr\}\nonumber
\\
&=&\min \biggl\{ n\dvtx  \frac{\rho_i(n)}{1-\rho_i(n)} \ge\frac
{1-{L}^{-1}}{{L}^{-1}} \biggr\}\nonumber
\\
&=&\min \biggl\{ n\dvtx  \log\frac{\rho_i(n)}{1-\rho_i(n)} - \log\frac
{\tilde{\rho}}{1-\tilde{\rho}} \ge
\log\frac{1-{L}^{-1}}{{L}^{-1}} - \log\frac{\tilde{\rho}}{1-\tilde{\rho}} \biggr\}
\\
&=&\min \biggl\{ n\dvtx  U_n \ge\log\frac{1-{L}^{-1}}{{L}^{-1}} - \log
\frac{\tilde{\rho}}{1-\tilde{\rho}} \biggr\}\nonumber
\\
&\le&\min \{ n\dvtx  U_n \ge\log{L} \}.\nonumber
\end{eqnarray}
The assertion of the theorem follows from~(\ref{tauredef}) and the
following lemma.
%which is a slight generalization of Lemma~6 in~\cite{Burnashev76}.

%
%le4 #&#
\begin{lemma}
\label{BurnashevUB} Consider the sequence $\{U_n\}$, $n=0,1,\ldots$
%$U_n:= \log\frac{\rho_i(n)}{1-\rho_i(n)} - \log\frac{\tilde{\rho}}{1-
defined in~(\ref{UDef}), and assume there exist positive constants $K_1
\le K_2 \le K_3$ such that
\begin{eqnarray*}
\mathbb{E} [ U_{n+1} | \mathcal{F}_n, \theta=i] &\ge&
U_n +K_1\qquad \mbox{if }U_n < 0,
\\
 \mathbb{E} [ U_{n+1} | \mathcal{F}_n, \theta=i] &\ge&
U_n +K_2\qquad \mbox{if }
U_n \ge0,
\\
\llvert U_{n+1} - U_n \rrvert &\le& K_3.
\end{eqnarray*}
Consider the stopping time $\upsilon= \min\{ n\dvtx  U_n \ge B \}$,
$B>[U_0]^+$. Then we have %the inequality
\[
\mathbb{E} [\upsilon|\theta=i] \le\frac{B - U_0}{K_2} + U_0 {\mathbf
{1}}_{\{U_0 <0\}} \biggl(\frac{1}{K_2} - \frac{1}{K_1} \biggr)+
\frac{K_3+\log e}{K_1}.
\]
\end{lemma}

The proof of Lemma~\ref{BurnashevUB} is provided in the supplemental
article~\cite{Naghshvar-SuppA}, Section~6. %\ref{supp:subm}.

%Assume that the sequence $\{\zeta_n\}$, $n=0,1,\ldots$ forms a
%submartingale with respect to a filtration $\{\mathcal{F}_n\}$.
%Furthermore, assume there exist positive constants $K_1 \le K_2 \le
%K_3$ such that
%& \mathbb{E} [ \zeta_{n+1} | \mathcal{F}_n ] \ge\zeta_n +K_1
%& \mathbb{E} [ \zeta_{n+1} | \mathcal{F}_n ] \ge\zeta_n +K_2
%& \left| \zeta_{n+1} - \zeta_n \right| \le K_3.
%Consider the stopping time $\tau_B = \min\{ n\dvtx  \zeta_n \ge B \}$,
%$B>0$. Then we have the inequality
%$$\mathbb{E} [\tau_B] \le\frac{B - \zeta_0}{K_2} + \zeta_0 {\mathbf{1}}_{
%3 \frac{K_3^2}{K_1 K_2}.$$

%In particular, let
%K_1&:=\max_{\bolds{\lambda} \in\mathbb{P}(
%K_{2i}&:=\max_{\bolds{\lambda} \in\mathbb{P}(
%q^{a}_j),\\
%K_3&:=\max_{i,j\in\Omega_M} \max_{a\in\mathcal{A}_M}
%}
%
In particular, from (C1)--(C3) and Lemma~\ref{BurnashevUB}, we have
\begin{eqnarray*}
&& \rho_i \mathbb{E} [
\tau_i | \theta=i]
\\
&&\qquad \le\rho_i \biggl( \frac{\log{L}- [\log (\rho
_{i}/(1-\rho_{i}))-\log(\tilde{\rho}/(1-\tilde{\rho}))
]^+}{D_{\bolds{\eta}_i}(M)}
\\
&&\hspace*{48pt}
{}+ \frac{ [\log ((1-\rho_{i})/\rho_{i})+\log(\tilde{\rho}/(1-\tilde{\rho})) ]^+
+ \xi_M+\log e}{I_2(M)} \biggr)
\\
&&\qquad \le\rho_i \frac{\log{L}}{D_{\bolds{\eta}_i}(M)} + \rho_i
\frac{\log (1/\rho_{i}) + \log(\tilde{\rho
}/(1-\tilde{\rho})) + \xi_M + \log e}{I_2(M)}. %&\le\rho_i \hspace*{-.03in} \left(\hspace*{-.02in}\frac{\log\L}{D_{
%%\nonumber
%%&\le\rho_i \frac{\log\L}{D_{\bolds{\eta}_i}(M)} +
%%\rho_i \frac{\left(\log\frac{1}{\rho_{i}} + \log\frac{\tilde{\rho}}{1-
%&\le\rho_i \frac{\log\L}{D_{\bolds{\eta}_i}(M)} +
\end{eqnarray*}
%
%Now from~\eqref{UBVpi2},~\eqref{UBK}, and the fact that $
%have the assertion of the theorem.
This inequality together with (\ref{UBVpi2}) and the fact that
$\sum_{i=1}^M \rho_i \log\frac{1}{\rho_{i}} = H(\bolds{\rho})$ implies
the assertion of the theorem:
%e51 #&#
\begin{eqnarray}
\quad V^*(\bolds{\rho}) &\le& V_{\tilde{\pi}_2}(\bolds{\rho})
\nonumber\\[-8pt]\label{UB2a} \\[-8pt]
&\le& \frac{H(\bolds{\rho})+\log (\tilde{\rho}/(1-\tilde
{\rho}))+\xi_M+\log e}{I_2(M)} + \sum_{i=1}^M
\rho_i \frac{\log{L}}{D_{\bolds{\eta}_i}(M)} + 1.\nonumber
\end{eqnarray}

\begin{remarks}
\label{remUB2b}
%It is shown in the supplemental article \cite[Section
%?]{Naghshvar-SuppA} that for large values of
For large values of $H(\bolds{\rho})$ and $\tilde{\rho}$ and when
$I_{\bolds{\eta}_0}(M) > I_{\bolds{\eta},\tilde{\rho}}(M)$, the upper
bound (\ref{UB2a}) can be tightened as follows (see Section 7 in \cite{Naghshvar-SuppA} for the proof):
%e52 #&#
\begin{eqnarray}
V_{\tilde{\pi}_2}(\bolds{\rho}) &\le& \frac{H(\bolds{\rho}) + \log(\tilde{\rho}/(1-\tilde
{\rho})) + \xi_M}{I_{\bolds{\eta}_0}(M)} +
\sum_{i=1}^M \rho_i\frac{\log{L}}{D_{\bolds{\eta}_i}(M)}
\nonumber\\[-8pt]\label{UB2b} \\[-8pt]
&&{}+ \frac{(1-\tilde{\rho})\log M + (2-\tilde{\rho})\xi_M + 4 + \log
e}{I_{\bolds{\eta},\tilde{\rho}}(M)} +1.\nonumber
\end{eqnarray}
\end{remarks}
\end{appendix}

\section*{Acknowledgments}
%We are grateful to two anonymous reviewers, the Associate
%Editor and Editor for their helpful comments which led to considerable
%improvements
%in the paper.
%%
%We would like to thank the Editor, an Associate Editor
%and two anonymous referees for insightful comments and suggestions
%which significantly
%improved our paper.
%%
%We would like to thank an Associate Editor and two
%anonymous referees for their insightful and constructive comments,
%which greatly
%improve the presentation of this paper.
%%
%The authors would like to thank the Associate Editor and
%the reviewers for their constructive comments, which substantially
%improve the
%paper.
%%
%We also thank the Editor, Associate Editor and two referees for their
%input which improved the paper.
%
We would like to thank Todd Coleman, Young-Han Kim, Bar{\i}\c{s}
Nakibo\u{g}lu, Yury Polyanskiy,
Maxim Raginsky, Sergio Verd\'u, Mich\`ele Wigger and Angela Yu for
valuable discussions and suggestions.
We are also grateful to the Editor, the Associate Editor and two
reviewers for their constructive comments. This work was done while Mohammad Naghshvar was with the Department of Electrical and Computer Engineering,
University of California, San Diego, La Jolla, CA 92093,~USA.
%which significantly helped to improve the paper.
%We greatly appreciate the constructive comments from the edito and the
%anonymous reviewers.

% AOS,AOAS: If there are supplements please fill:
%
\begin{supplement}\label{SuppA} %[id=suppA]
\stitle{Technical proofs}
\slink[doi]{10.1214/13-AOS1144SUPP}
\sdatatype{.pdf}
\sfilename{aos1144\_supp.pdf}
\sdescription{For the interest of space, we only provided the proofs
of the theorems in this paper.
Proofs of the propositions, lemmas, corollaries and technical claims
are provided in the supplemental article.} %Supplement~A.}
\end{supplement}

%suskaldyti doi

% imsref loaded by linak, 2013-11-13 18:04:50
%

\printaddresses

\end{document}